\documentclass[plb,twocolumn,superscriptaddress,nofootinbib]{revtex4-1}
\usepackage{epsfig}
\usepackage{graphicx}
\usepackage{palatino}
\usepackage[english]{babel}
\usepackage{hyphenat}
\usepackage{amsmath}
\usepackage{amssymb}
\usepackage{slashed}
\usepackage{epstopdf}
\usepackage{xcolor}
\usepackage{booktabs}
\usepackage[toc,page]{appendix}
\definecolor{lcolor}{rgb}{0.,0.0,0.}
\definecolor{citcolor}{rgb}{0,0.,0.5}
\usepackage[breaklinks,colorlinks,urlcolor=blue,citecolor=blue,linkcolor=blue]{hyperref}
\usepackage{multirow}
\usepackage{ltablex}
\usepackage{cancel}
\usepackage{ulem}
\newcommand{\Pythia}{\textsc{Pythia}}
\newcommand{\Latekt}{Late-$k_t$}
\newcommand{\beq}{\begin{equation}}
\newcommand{\eeq}{\end{equation}}
\newcommand{\bea}{\begin{eqnarray}}
\newcommand{\eea}{\end{eqnarray}}

\def\dd{{\rm d}}

\newcommand{\bem}{\begin{multline}}
\newcommand{\eem}{\end{multline}}
\newcommand{\beg}{\begin{gather}}
\newcommand{\eeg}{\end{gather}}

\newcommand{\ben}{\begin{eqnarray*}}
\newcommand{\een}{\end{eqnarray*}}

\def\and{\quad\text{and}\quad}

\def\cut{\text{cut}}

\newcommand{\ktcut}{k_{t,\rm{cut}}}
\newcommand{\barktcut}{\bar{k}_{t,\rm{cut}}}

\begin{document}

\title{Dead-cone searches in heavy-ion collisions using the jet tree}

\author{Leticia Cunqueiro}
\email[]{leticia.cunqueiromendez@uniroma1.it}
\affiliation{Sapienza Universit\`a di Roma \& INFN, Piazzale Aldo Moro 5, Roma 00185, Italy }
\author{Davide Napoletano}
\email[]{davide.napoletano@unimib.it}
\affiliation{Universit\`a degli Studi di Milano-Bicocca \& INFN, Piazza della Scienza 3, Milano 20126, Italy}
\author{Alba Soto-Ontoso}
\email[]{alba.soto.ontoso@cern.ch}
\affiliation{Theoretical Physics Department, CERN, 1211 Geneva 23, Switzerland}

\begin{abstract}
We explore the possibility of using the dead cone of heavy quarks as a region of
the Lund plane where medium-induced gluon radiation can be isolated and
characterised. The filling of the dead cone by medium-induced gluons is expected
to be the result of the interplay between the minimum angle of such radiation
due to transverse momentum broadening and the dead-cone angle. Since the measurement of a fully corrected Lund plane in heavy-ion collisions is currently challenging, we propose to use jet grooming techniques to identify a particular splitting in the jet tree that is both perturbative and sensitive to the dead-cone effect. To that end, we propose a new jet substructure groomer, dubbed \Latekt, that selects the most collinear splitting in a QCD jet above a certain transverse momentum cutoff $\ktcut$. The role of $\ktcut$ is to guarantee perturbative splittings, while selecting the most collinear splitting enhances the sensitivity to mass effects.
 As a proof of concept, we study the angular distribution of the splitting tagged by \Latekt~both analytically and with Monte Carlo simulations. First, we derive the logarithmic resummation structure in vacuum and demonstrate its capability to distinguish between inclusive and heavy-flavoured jets. Next, we extend the calculation for in-medium jets and show that medium-induced emissions lead to an enhancement of collinear emissions below the dead cone angle. Numerically, we demonstrate an excellent resilience of \Latekt~against uncorrelated thermal background, thus confirming this observable as a potential candidate to unveil medium dynamics around the dead cone regime.
\end{abstract}

\maketitle
\section{Introduction}
One of the fundamental properties of Quantum Chromodynamics (QCD) is the
depletion of collinear radiation when the emitter is a massive
quark~\cite{Marchesini:1989yk,Dokshitzer:1991fd}. More concretely, radiation from an energetic,
massive quark is strongly suppressed for angles smaller than the so-called
dead cone angle, i.e.
\beq
\label{eq:deadcone-def}
\theta_{0}\equiv \frac{m_{Q}}{E},
\eeq
where $m_Q$ denotes the quark mass and $E$ its energy.
Thus, while the probability for an emission at an angle $\theta$
diverges in the collinear limit as $\propto 1/\theta$ for light quarks,
the probability of radiating a gluon off a massive quark is collinear finite since
it scales as $\propto \theta^3/(\theta^2+\theta_0^2)^2$.

This cornerstone property of the strong force has recently been experimentally
proven for charm quarks by the ALICE collaboration~\cite{ALICE:2021aqk}.
The measurement was based on the method proposed in
Ref.~\cite{Cunqueiro:2018jbh} and consisted in the use of iterative declustering
to build the primary Lund plane~\cite{Dreyer:2018nbf,Lifson:2020gua} of jets
with a fully-reconstructed D meson as a constituent.
The iterative declustering allows to select certain nodes of the angular-ordered
jet tree that are a proxy for a splitting of the heavy-flavour quark $Q$ and for
which the subleading prong represents the gluon emission in the $Q \to Qg$
process.
The angular distribution of such splittings was compared to similarly obtained
splittings in inclusive jets and a clear suppression of small-angles was
observed for D-tagged jets.
This suppression was found to qualitatively satisfy Eq.~\eqref{eq:deadcone-def}
when using the splitting energy as a proxy for the heavy-quark energy.
To follow up on this, ALICE has presented a prospect for the measurement of the
dead cone in $B$-jets using similar techniques in Run3~\cite{ALICE:2020fuk}.
It is important to notice that for a precise measurement of the dead cone in
high-energy jets, the detector needs
to resolve small angles. The recent measurement of the track-based Lund plane by ATLAS~\cite{ATLAS:2020bbn} demonstrates the experimental low-angle reach by reporting splitting angles down to 0.005 radians, which is compatible with the size of the pixel pitch of their silicon tracker.

The successful measurement performed by ALICE lies on three essential ingredients.
First, the ability to penetrate the jet tree down to the splittings at the
smallest angles for which quark mass effects are enhanced.
Second, the ability to suppress hadronisation effects that fill the dead cone,
by imposing a minimum requirement on the hardness -- $k_{t}$ -- of the
splittings, and third, the ability to fully reconstruct heavy-flavour hadrons.
Without a full reconstruction of the heavy-flavour hadron, the decay products
interfere with the reclustering process and distort the jet tree creating extra
splittings at small angles that darken the dead cone.

The possibility of producing a similar measurement in a heavy-ion
environment may represent a unique opportunity to study a region of phase-space
dominated by medium-induced emissions, i.e. triggered by the interactions
of the hard propagating parton and the medium. In fact, a distinct feature of jets embedded in a
Quark-Gluon Plasma (QGP) is that the branching probability for
medium induced emissions is, in contrast to the
vacuum case, collinear finite even for massless quarks.
This shielding of the collinear divergence
is due to the transverse momentum acquired by the radiation during its
formation time via elastic scatterings with the QGP.
However, if the propagating quark is massive
this effect is suppressed~\cite{Armesto:2003jh,Dokshitzer:2001zm,Djordjevic:2003zk,Zhang:2003wk,Aurenche:2009dj} due to the formation time of an emission off a massive quark being
reduced relative to the emitter being massless
\beq
\label{eq:tf-vac}
t_f^{\mathrm{massive}} \equiv \frac{2}{\omega(\theta^2+\theta_0^2)}\,
> \frac{2}{\omega\,\theta^2}\equiv t_f^{\mathrm{massless}} \,,
\eeq
where $\omega$ is the frequency of the gluon.
Then, one can identify two competing mechanisms acting on
the medium-induced spectrum of massive quarks.
On the one hand, the dead cone suppresses collinear radiation just as in vacuum.
On the other hand, the reduction of LPM-like interference effects leads to a
higher intensity of the medium induced spectrum even for angles
smaller than the dead cone.
Thus, an excess of collinear
emissions at angles $\theta<\theta_0$ for heavy-flavoured tagged jets in
heavy-ion collisions with respect to proton-proton would serve to isolate
medium-induced dynamics.

It is important to keep in mind that the way this
comparison is made between heavy-ions and $pp$ results is critical for a
robust interpretation of the data uniquely in terms of dead-cone effects.
As an example, another mechanism that can generate an excess of collinear
splittings, even at the inclusive level, is colour decoherence dynamics.
That is, energy loss is weaker for unresolved splittings,
i.e. branchings with opening angles smaller than the QGP resolution
angle,
$\theta_c$~\cite{Mehtar-Tani:2010ebp,Mehtar-Tani:2011hma,Casalderrey-Solana:2011ule}.
Since the jet spectrum is steeply falling, any jet analysis that imposes a
$p_t$ selection automatically favour jets that lost little energy,
i.e. narrower jets.
As a consequence, any comparison between heavy-flavour and inclusive jets in
heavy-ions and $pp$ is subject to both dead-cone and colour coherence effects.

To be able to perform this measurement in medium,
the aim of this work is to propose an observable capable of satisfying the three principles
that made the ALICE measurement in vacuum possible, while, at the same
time, maximising the medium-induced region and being relatively insensitive to
colour coherence effects.
To achieve this, as a first step, we introduce a new grooming technique -- dubbed \Latekt~ --
designed to be naturally sensitive to small-angle radiation, while at the same
time being relatively insensitive to hadronisation effects and/or underlying event.

In order to highlight the properties of the newly proposed \Latekt~groomer, we compare it to other well-known groomers. Take as an example Soft Drop (SD)~\cite{Dasgupta:2013ihk,Larkoski:2014wba},
which tags the first emission satisfying the condition
$z>z_{\rm cut}\theta^\beta$, with $\beta$ a free parameter, grooming away softer
emissions. As studied in~\cite{Mulligan:2020tim}, in the case of heavy-ion
collisions, one needs a relatively large $z_{\rm cut}\sim 0.2$ and $\beta=0$ to reduce the
sensitivity to the uncorrelated thermal background. This choice of SD parameters
effectively corresponds to a typical angle of $\theta \sim 0.2 \gg \theta_0$,
for jets of radius $R=0.2$~\cite{ALargeIonColliderExperiment:2021mqf}. In addition to
being sensitive to large angle emissions, the $k_t$ of an SD
emission can be arbitrary small and, as a consequence, the small angle regime
$\theta_g\sim\theta_0$ is affected by sizeable non-perturbative corrections, especially for
charm-initiated jets. A logic solution to avoid this problem is to choose
$\beta=1$, which would correspond in practice to a $k_t$ cut.
In the following we refer to this option as \textit{Early-$k_t$}, and one can see that
this choice also gives rise to potential issues. Indeed, one then becomes sensitive to
emissions at all angles (preferentially around
$\theta\sim R$ for a not too low value of $\ktcut$), thus
potentially undermining the discriminating power between light and massive quarks.

Building up on these ideas, with \Latekt~we identify the most collinear
splitting above a certain transverse momentum threshold $\ktcut$ and remove
all emissions that occur at larger angles, i.e. splittings with
$\theta>\theta_\ell$.
As we show in the rest of this manuscript, the main advantages of this groomer
is the presence of a $\ktcut$, which ensures that mainly perturbative
splittings prevail while keeping good purities in the presence of the heavy-ion background.
Our proposed observable to search for dead cone effects is then the angular
distribution of the tagged \Latekt~splitting.

Grooming techniques fall in the realm of jet-substructure calculations, which
have seen wide use in both $pp$ and heavy-ion collisions -- see
Refs.~\cite{Marzani:2019hun,Larkoski:2017jix,Cunqueiro:2021wls} and references
therein -- as they provide powerful theoretical tools for perturbative-QCD
calculations as well as for robust theory-data comparison -- see
Refs.~\cite{ATLAS:2017zda,ATLAS:2019mgf,Tripathee:2017ybi,STAR:2020ejj,ALICE:2022hyz,Marzani:2017kqd,Kang:2019prh,Cal:2021fla,Frye:2016aiz,Kang:2018vgn}
as examples.
However, most of the recent progress in this field has been devoted to jets
initiated by massless quarks. Progress on heavy-flavour jet substructure has been relatively
slow and only a few examples of resummed
calculations for this class of observables exist in the literature either
in vacuum~\cite{Makris:2018npl,Lee:2019lge,Craft:2022kdo} or in medium~\cite{Li:2017wwc}.
In fact, the first NLL calculation of a heavy-flavoured jet substructure
observable in $pp$ collisions appeared only very recently~\cite{Craft:2022kdo}.
Such theoretical developments are timely and relevant both for the upcoming run
of the LHC, as well as for future lepton colliders,
where heavy flavour jet production will play an even more important role for
precision studies.
It is worth pointing out that the authors of Ref.~\cite{Craft:2022kdo} proposed
to use energy correlators to expose the dead cone effect in proton-proton
collisions. Another idea proposed in the literature is to use
boosted top-quark jets~\cite{Maltoni:2016ays}.
Compared to both of these works, we take an alternative approach and opt to
minimally extend the clustering-tree-based observable that lead to the
dead-cone discovery in $pp$ to a heavy-ion environment.

The rest of this work studies the properties of the $\theta_\ell$-distribution as
defined by \Latekt, and its comparison to other groomers.
We first investigate, using well-known analytic resummation techniques, its behaviour
both for the case of vacuum emissions and for in medium propagation.
Second, we show the resilience of this observable to non-perturbative dynamics
and to thermal background effects in heavy-ion collisions by means of realistic
Monte Carlo (MC) simulations. Last, we report our conclusions.

\section{Late-$k_t$}
\label{sec:latekt}

\begin{figure*}[t]
    \centering
    \includegraphics[width=0.99\textwidth]{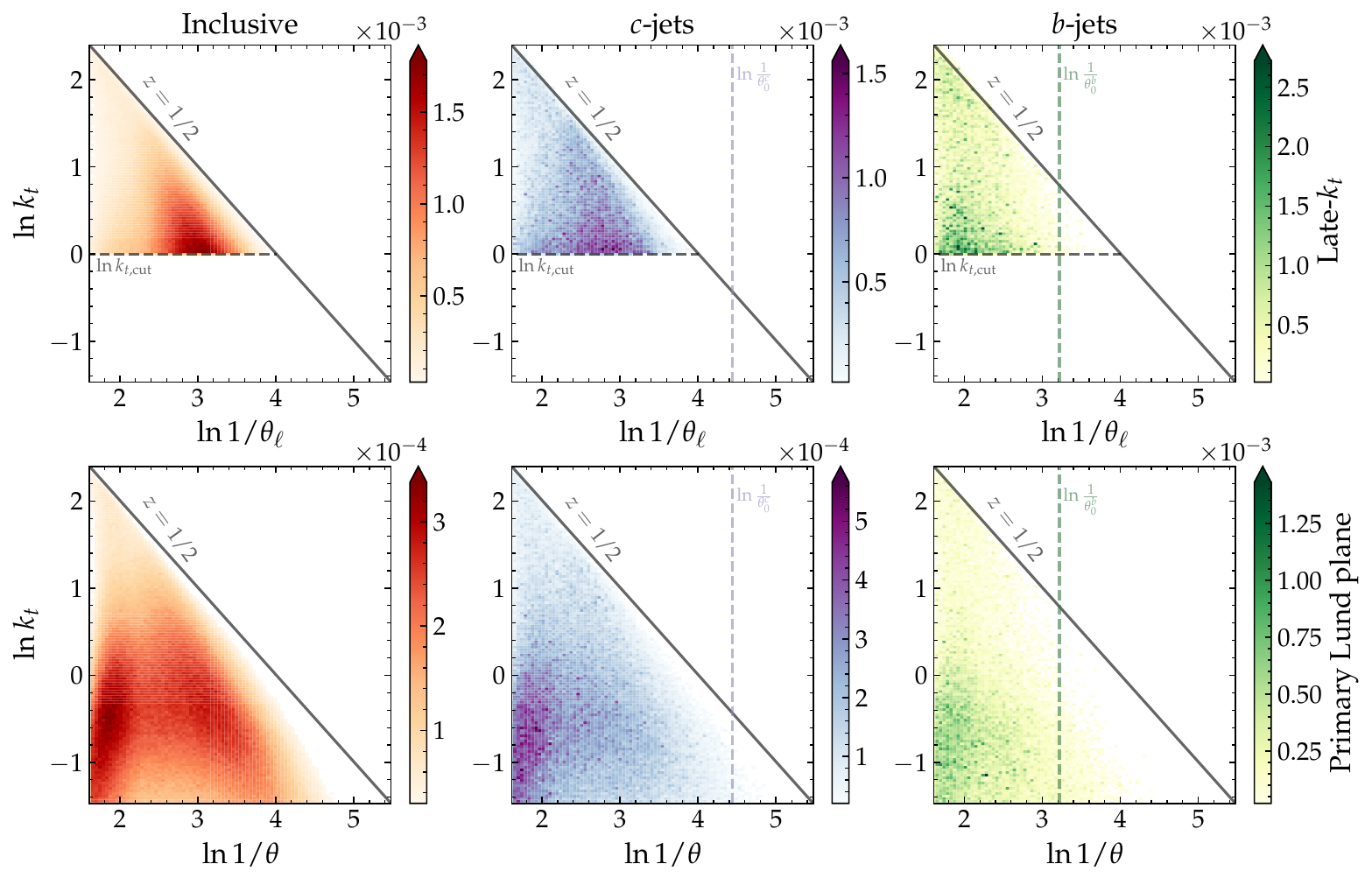}
    \caption{Top: Lund plane density of the splitting tagged by \Latekt~for inclusive jets (left), $c$-jets
      (middle) and $b$-jets (right) obtained with \Pythia8 at hadron level and including underlying event. The vertical, dashed, black lines indicate the dead cone angle of the first splitting in the massive cases.]
      Bottom: same as top panel but including all splittings along the primary branch.}
    \label{fig:lp-pythia}
\end{figure*}

In this section we define and study the main properties of the \Latekt~groomer.
It is based, similarly to many other groomers and taggers, on a
declustering procedure. In particular,
given a jet, obtained with any available jet clustering, we start by
re-clustering it with a Cambridge-Aachen
(C/A)~\cite{Dokshitzer:1997in,Wobisch:1998wt} and identify all $n$-splittings along the primary branch
whose transverse momentum is above a given threshold, i.e. $k_t>\ktcut$.
Then, among this list of splittings we select the one with the smallest angle,
i.e. $\theta_\ell \equiv \min \lbrace \theta_1, \theta_2, \ldots, \theta_n
\rbrace$. All emissions occurring at larger angles are groomed away.
In practice, this corresponds to the following algorithmic procedure:
\begin{enumerate}
\item \label{item:step1}
Undo the last clustering step in the angular ordered jet to produce two pseudojets with momenta $p_a$ and $p_b$.
\item \label{item:step2} Calculate the relative $k_t$ of the pair
\begin{equation}
    k_t = \text{min}(p_{t,a},p_{t,b})\Delta_{ab},
\end{equation}
with $\Delta^2_{ab}$ being the relative distance in the rapidity-azimuth plane,
i.e. $\Delta^2_{ab}=(y_a-y_b)^2+(\phi_a-\phi_b)^2$. Note that this definition coincides with the Lund-$k_t$
variable introduced in Ref.~\cite{Dreyer:2018nbf}.
\item\label{item:step3}
Store the value of $\Delta_{ab}$ only if $k_t>\ktcut$, and repeat from step~\ref{item:step1} following the hardest branch.
\item\label{item:step4}
Finally, find the minimal value in the list of $\Delta_{ab}$ values and drop all branches at larger angles, that is, prior in the C/A sequence.
\end{enumerate}
For heavy-flavour tagging we also impose, at each declustering step, that
the leading prong must contain at least one heavy-flavour hadron within its
constituents as it was already done in the successful measurement of the
dead cone in $pp$ collision~\cite{ALICE:2021aqk}.

As a visual example of the effect of applying this groomer, we report in Fig.~\ref{fig:lp-pythia}
the Lund plane density both for all splittings along the primary branch, and for just the splitting tagged by
\Latekt.
To obtain this plot we produce dijet events at $\sqrt s=5.02$~TeV with
\Pythia8~\cite{Bierlich:2022pfr}, disabling $B$ and
$D$ decays. We then reconstruct jets using the anti-$k_t$~\cite{Cacciari:2008gp} algorithm -- as implemented in
FastJet~\cite{Cacciari:2011ma} -- with a jet radius of $R=0.2$. Further, we
label $c$- or $b$-jets those jets having at least one $D$ or $B$ hadron among
their constituents, and any other jet is labelled as \textit{inclusive}.
Lastly, we apply the Cambridge/Aachen
algorithm to re-cluster our jets.
In the case of Late-$k_t$ we follow the algorithm explained above
setting $\ktcut = 1$ GeV, while for the primary Lund plane we record
the $k_t$ and $\theta$ values of all splittings in the leading branch as
explained in Ref.~\cite{Dreyer:2018nbf}.

The impact of applying the \Latekt~groomer depends on the jet flavour,
as it is clearly visible in Fig.~\ref{fig:lp-pythia}. In the inclusive case, its effect
is to cut low transverse momentum emissions, and give more
relative importance to emissions at smaller angles. This aspect is crucial for dead cone
searches since it is precisely in the collinear regime where we expect largest differences between
light and heavy emitters. Since we select $\ktcut\sim m_{c}$,
part of the dead-cone region for charm jets is groomed away. As a consequence,
the phase-space of inclusive and charm jets are more alike after grooming
than in the plain case. However, for bottom initiated jets the hierarchy
$\ktcut < m_{b}$ allows to preserve the sensitivity to the dead cone
and thus the differences between $b$-jets and inclusive are enhanced after grooming.
Since a full Lund plane measurement in heavy-ion collisions is challenging and not yet attempted experimentally,
we dedicate the remainder of this section to study the $k_t$-integrated version of
Fig.~\ref{fig:lp-pythia}.

\subsection{Analytic estimates}
\label{sec:analytics}
To gain some analytic insight on the sensitivity of
the $\theta_\ell$-distribution to the dead-cone angle both for vacuum and
in-medium jets, we now present analytic calculations for the opening angle, $\theta_\ell$,
distribution of the splitting tagged by Late-$k_t$.
Note that while the formulae presented in this work are focused on the angle
of the splitting, they can readily be extended
to other substructure observables, such as the splitting mass, or its relative
transverse momentum.
We stress that the aim of the following calculation is to discuss the qualitative
features of the $\theta_{\ell}$-distribution and thus we target
modified leading-logarithmic accuracy,
which includes leading hard-collinear effects on
top of purely soft-and-collinear ones, and suffices for this purpose.
We additionally note that, for heavy-quarks, the strictly collinear limit is shielded by the mass
of the quark, and the relevant limit is the so-called quasi-collinear
limit, as defined for example in Ref.~\cite{Catani:2002hc}.
In the following we use the two words interchangeably where no confusion arises.
Further, we consider the jet energy to be identical to the jet transverse
momentum.
Since we do not consider energy degradation, this also coincides with the
energy of the splitter and, as such, the dead cone angle is fixed.
Note that this is not the case in a realistic parton shower, as explained
in Sec.~\ref{sec:experimental}.

\subsubsection{Vacuum}
\label{sec:vacuum}
 \begin{figure*}[t]
    \centering
    \includegraphics[width=0.99\textwidth]{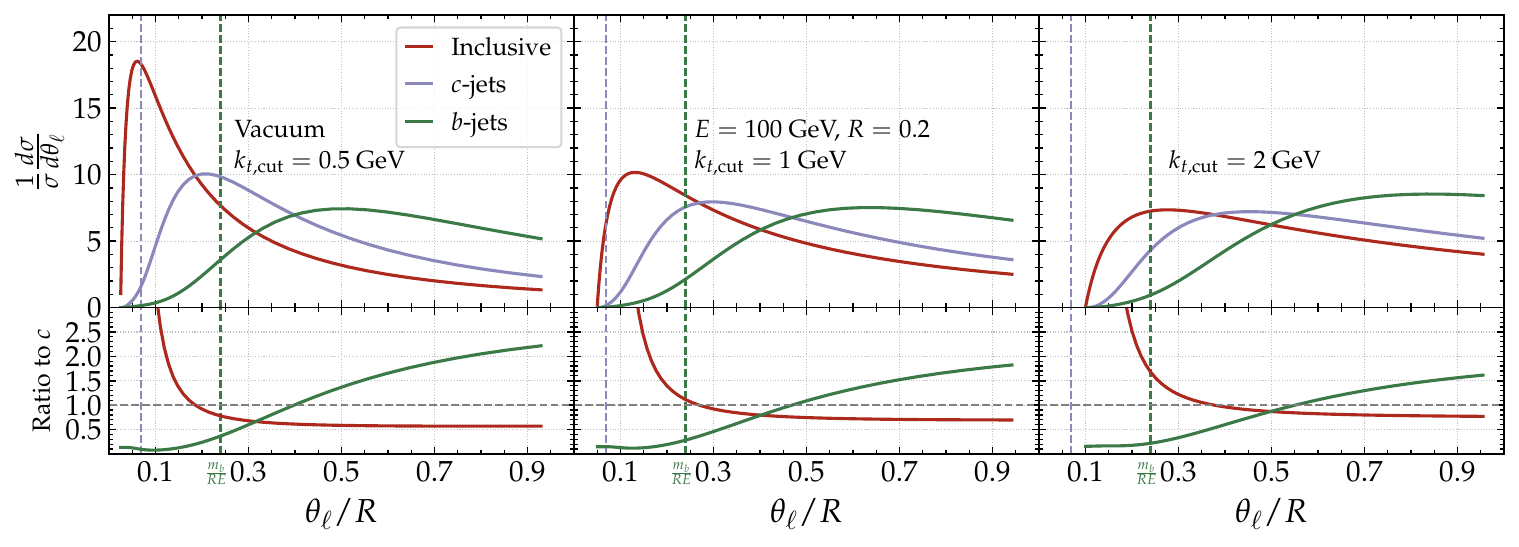}
    \caption{Angular distribution of the splitting tagged by Late-$k_t$ for
      inclusive jets (red), $c$-jets (purple) and $b$-jets (green) for three
      different values of the $\ktcut$ [GeV]: 0.5 (left), 1 (middle) and 2 (right).
      The bottom panel displays the ratio to charm jets and the vertical
      dashed lines denote the positions of the dead cone angle of the first splitting of the jet tree.}
    \label{fig:thg-vacuum}
\end{figure*}

We start by considering a vacuum jet and calculate the cumulative
distribution for $\theta_\ell$, that is, the
probability to tag a splitting with an angle below $\theta_\ell$. This is given
by
\begin{equation}
\Sigma(\theta_\ell) = \int_0^{\theta_\ell} \dd \theta'\frac{1}{\sigma}\frac{\dd \sigma}{\dd \theta'}\,.
\end{equation}
To practically compute this, we consider the independent emission of $n$
collinear partons, either real or virtual, within a jet of radius $R$. Our calculation is
based on the modified leading-logarithmic approximation that assumes angular
ordering, i.e. $R\gg\theta_1\gg \theta_2 \gg \ldots \gg \theta_n$, while the energy fractions of the emissions, $z_i$, are not strongly ordered~\cite{Marzani:2019hun}.
The cumulative distribution for an $i=(q,g)$-initiated jet
then reads
\begin{equation}
\label{eq:master}
  \begin{split}
    \Sigma(\theta_\ell) & = \sum_{n=1}^\infty\frac{1}{n!}
    \prod_{m=1}^n\int \dd \theta_m
 \int\dd z_m\,
    \frac{\alpha_s(k_{t,m})}{2\,\pi}\mathcal{P}_i(z_m,\theta_m) \\
  & \times \Theta(\barktcut-z_m\theta_m)\Theta(\theta_\ell-\theta_m)\, \\
   & -  \sum_{n=1}^\infty\frac{1}{n!}
    \prod_{m=1}^n\int \dd \theta_m
    \int\dd z_m\,
    \frac{\alpha_s(k_{t,m})}{2\,\pi}\mathcal{P}_i(z_m,\theta_m) \\
    & \times\Theta(\theta_\ell-\theta_m)
  \end{split}
\end{equation}
where $\alpha_s(k_t)$ is the strong coupling constant, $\barktcut =
\ktcut/p_{t}$, and we denote by $\mathcal {P}$ the branching kernel that we
specify later in this section.

The physical interpretation of
Eq.~\eqref{eq:master} is the following. Since we define $\theta_\ell$ as the smallest
angle that satisfies the \Latekt~condition, we need to ensure that all other
real emissions at smaller angles do not pass the $\ktcut$. This is encoded in the
first two lines of Eq.~\eqref{eq:master}. The last two lines correspond to the
virtual corrections that have opposite sign, following the KLN theorem, and no
$\ktcut$ condition since a virtual emission can never trigger \Latekt.
By performing the sums in Eq.~\eqref{eq:master}, we obtain the cumulative distribution or, analogously, the Sudakov form factor
\begin{equation}
\label{eq:vac-sigma-thetag}
\begin{split}
    \ln\Sigma(\theta_\ell) = &-\int_{0}^{\theta_\ell} \dd \theta
 \int_0^{1}\dd z
    \frac{\alpha_s(k_t)}{2\,\pi}\mathcal{P}_i(z,\theta) \\
    &\times \Theta(z\theta p_{t} - \ktcut).
    \end{split}
\end{equation}

Eq.~\eqref{eq:vac-sigma-thetag} is reminiscent of the Sudakov form
factor for the regular Soft Drop tagger (see Eq.~(4.1) in
Ref.~\cite{Larkoski:2014wba}), and could, in principle, be similarly generalised
by performing the following replacement
\begin{equation}
  \nonumber
   \Theta(z\theta p_{t} - \ktcut) \rightarrow \Theta(z\theta^\beta p_{t} - x_{\cut})\,.
 \end{equation}
In doing so, it is important to stress that there is a fundamental difference
with respect to the regular Soft Drop tagger.
While Soft Drop
selects the first splitting that meets the tagging condition and therefore
vetoes all splitting at larger angles, i.e. $\int_{\theta_{\ell}}^R \dd \theta$, the
Late-$k_t$ procedure selects the most collinear splitting and consequently
vetoes all splittings at smaller angles.
As such, only for $\beta>0$ one is guaranteed to obtain a collinear finite result for
massless splitters. For massive splitters the collinear singularity is automatically shielded by their mass, and thus one have to instead be careful
that the value of the $\ktcut$ does not interfere with dead-cone dynamics.

Equipped with Eq.~\eqref{eq:vac-sigma-thetag}, we can construct the normalised
probability distribution for the $\theta_{\ell}$ distribution,
\begin{equation}
\label{eq:vac-thetag}
   \frac{1}{\sigma}\frac{\dd \sigma}{\dd \theta_{\ell}}=\frac{1}{1-\Sigma(R)}
 \int_0^{1}\dd z
    \mathcal{P}_i(z,\theta_\ell)\Theta(z\theta_\ell -
    \barktcut)\, \Sigma(\theta_\ell),
\end{equation}
where the $1-\Sigma(R)$ factor arises as a result of the maximum value of
$\theta_\ell$ being the jet radius. In order to evaluate the previous equation,
we have to specify the exact form of the branching kernels. They are given by
\beq
\label{eq:branch-kernel-vac}
{\cal {P}}_{i}(z,\theta) = \frac{\alpha_s(k_t)}{\pi}\frac{\theta}{\theta^2 + \Big(\frac{\theta_0}{1-z}\Big)^2} P_i(z,\theta),
\eeq
where $P_i$ are the unregularised Altarelli-Parisi splitting functions that read
\begin{align}
\label{eq:split-functions}
P_Q(z) &= C_F\left[\frac{1+(1-z)^2}{z} - \frac{2\theta_0^2}{z(1-z)\Big[\theta^2+\Big(\frac{\theta_0}{1-z}\Big)^2\Big]}\right]
\end{align}
and we have set the masses to zero in $g\to Q\bar Q$ splittings since, when
considering the inclusive case we treat it as massless.
Finally, within the modified leading logarithmic approximation (MLLA), we consider the
running of the strong coupling constant at 1-loop,
\begin{equation}
\label{eq:alphas-1loop}
\begin{split}
\alpha_s(k_t) &= \frac{\alpha_s}{1+2\beta_0\alpha_s\ln\frac{k_t}{Q}} \\
& = \alpha_s + 2\alpha_s^2\beta_0 \ln\frac{Q}{k_t}+\mathcal{O}(\alpha_s^3),
\end{split}
\end{equation}
with $\alpha_s\equiv\alpha_s(Q)$ and the hard scale chosen to be $Q=p_{t}R$.
In order to avoid the Landau pole, we freeze the coupling in the infrared
below $\ktcut$. With these choices for $\alpha_s$ and the branching kernels, all integrals in
Eq.~\eqref{eq:vac-thetag} can be done analytically. We consider three different settings: (i) inclusive,
where we set $\theta_0=0$ and weight the quark and gluon contributions by
their corresponding Born-level cross section extracted from \Pythia8 simulations,
(ii) charm jets, where we use $m_c =1.4$~GeV, and
(iii) bottom jets, where we use $m_b = 4.8$~GeV.

The results for the $\theta_\ell$-distribution are shown in
Fig.~\ref{fig:thg-vacuum} for jets with $E=100$ GeV, cone size $R=0.2$, and
different values of $\ktcut$. We observe that, by construction, the
\Latekt~tagger yields distributions peaked in the collinear regime and that the
distributions shift towards larger angles when increasing the initiator mass, as
desired. As we anticipated, the value of $\ktcut$ is critical for an
interpretation of the curves in terms of dead-cone dynamics, particularly for
$c$-jets. More concretely, we observe that for $\ktcut<m_c$ a clear ordering
exists in the peak position of the inclusive and $c$-jets curves, being the
inclusive curve peaked at smaller angles. However, when increasing the
transverse momentum threshold to $\ktcut=2$~GeV, the discriminating power of
\Latekt~between inclusive and charm jets is reduced. This is a
consequence of the artificial infrared cutoff $\theta=\ktcut/p_{t}$ mimicking
the charm dead cone angle. Regarding $b$-initiated jets, the difference with
respect to the massless benchmark shrinks when increasing the $\ktcut$, although
the effect is not as dramatic as for charm due to the bottom quark mass being
larger. On the other hand, as we discuss in Sec.~\ref{sec:experimental}, setting $\ktcut$ to very small values leads to a strong sensitivity of this observable to hadronisation corrections. Therefore, even if $\ktcut$ is a free parameter of the model, only a narrow window of values around 1 GeV maximise the sensitivity to the dead-cone angle while keeping non-perturbative corrections under control.

\subsubsection{Medium}
\label{sec:medium}
 \begin{figure*}[t]
    \centering
    \includegraphics[width=0.99\textwidth]{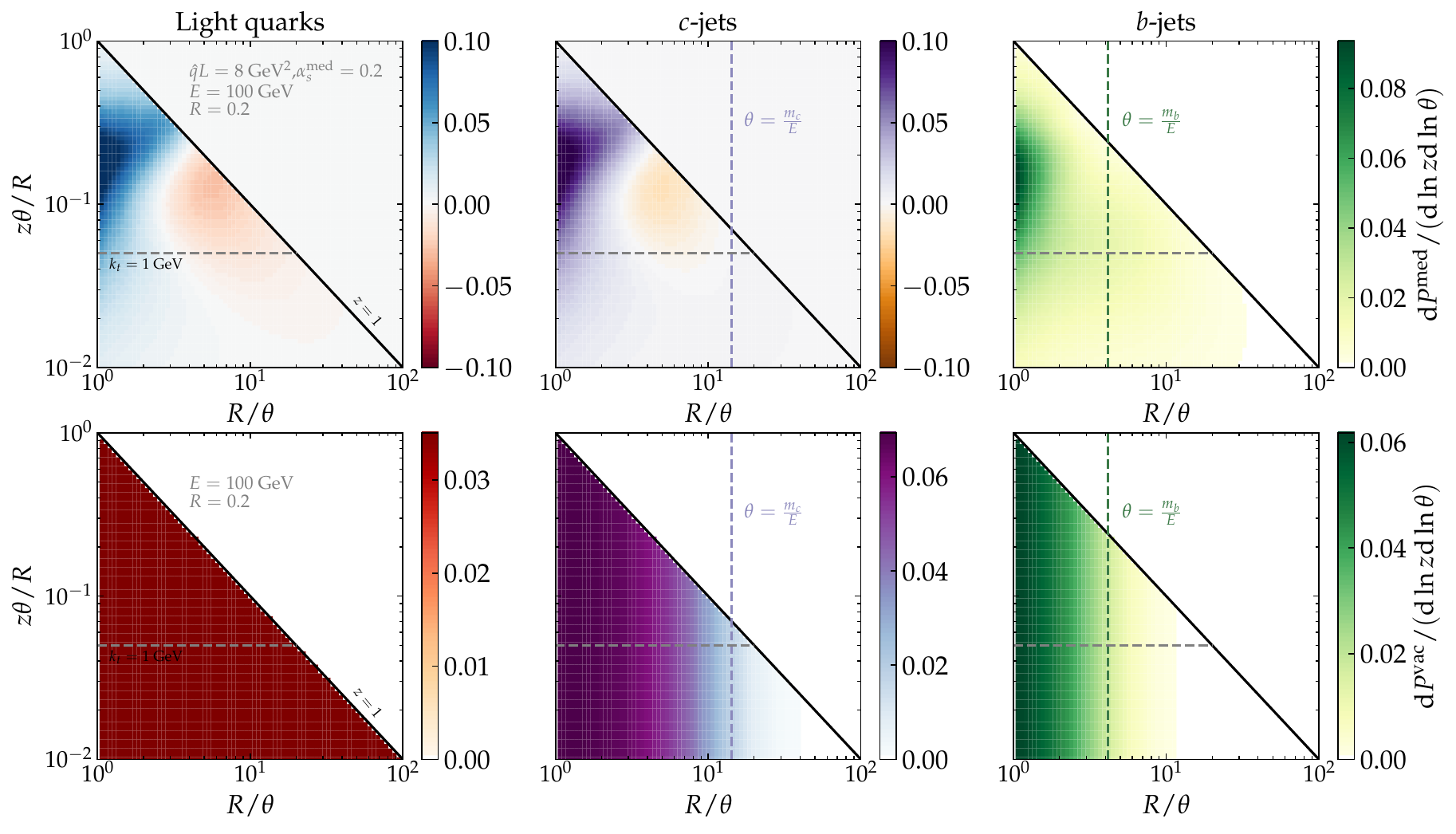}
    \caption{Top: Medium-induced Lund plane density for jets initiated by either a light (left), charm (middle) or bottom quark (right). As usual, in the heavy-flavoured cases we also draw the lines corresponding to dead cone angles of the first declustered splitting. Bottom: same as top panel but for vacuum emissions in the double-logarithmic approximation.}
    \label{fig:lp-medium}
\end{figure*}

In this section, we study the imprint of a hot, thermal-medium such as the
Quark-Gluon Plasma on the opening angle of the splitting tagged by \Latekt. As
was the case in other jet substructure
calculations~\cite{Mehtar-Tani:2016aco,Caucal:2021cfb,Pablos:2022mrx}, we
describe the interaction between the hard-propagating parton and the medium in
the limit of multiple, soft scatterings and leave the study of higher-order
corrections to this picture~\cite{Blok:2019uny} for future works. The medium is
considered to be a static brick of length $L$ and it's characterised by the
diffusion parameter $\hat q$. The characteristic scale of the medium is then
given by the product of the two and represents typical transverse momentum squared
acquired after propagating through the whole plasma. We explore values of $2<
\hat q L<8$ GeV$^2$ as was done in Ref.~\cite{Attems:2022otp}. Finally, we only
consider gluon emissions off quark legs and restrict the calculation to the soft
($\omega\ll E$) and collinear ($\theta^2\sim k_t^2/\omega^2\ll 1$) limit of QCD,
i.e. we achieve double-logarithmic accuracy, including power corrections
proportional to the dead-cone angle. In order to have meaningful comparisons
between vacumm and medium results, results presented in this section for vacuum
are obtained in the leading-logarithmic approximation, obtained by setting  the
splitting functions $P(z)\to 1/z$, and by fixing the strong coupling constant to
$\alpha_s(p_tR)$, thus removing hard-collinear effects.

We now consider the case in which the emission tagged by \Latekt~is
medium-induced. In this case, the vacuum branching kernel in
Eq.~\eqref{eq:vac-thetag} needs to be replaced by the medium-induced spectrum
for a gluon emission off a hard, massive quark leg. This was calculated almost
two decades ago in Ref.~\cite{Armesto:2003jh}, where it was realised that the
mass of the emitter introduces a dead-cone factor together with an additional
oscillatory phase in the path-integral. The spectrum was calculated both in the
multiple soft scattering approximation~\cite{Baier:1996sk,Baier:1996kr}, and
following an opacity expansion
approach~\cite{Wiedemann:2000za,Gyulassy:2000fs}.
Here we stick to the former for which the fully differential spectrum reads:
\begin{align}
\label{eq:spectrum-massive}
\omega &\frac{\dd^2 I}{\dd \omega \dd \kappa^2} = \frac{2\alpha^{\rm med}_s C_F}{\pi}\gamma\Bigg\lbrace\gamma {\text{Re}}\int_0^1 \dd t \int _t^1 \dd\bar t e^{ i M^2\gamma (t-\bar t)}  \nonumber \\
&\times  \exp\left[-\frac{\kappa^2}{4(D-i A B)}\right] \left[\frac{i A^3 B \kappa^2}{(D-i A B)^3} - \frac{4 A^2 D}{(D-i AB)^2} \right] \nonumber \\
& +  {\text{Re}} \int_0^1 \dd t e^{-i M^2 \gamma t}\left(\frac{-i\kappa^2}{\kappa^2+M^2}\right)\frac{1}{\cos^2(\Omega t)}\nonumber \\
&\times \exp\left[\frac{-i\kappa^2}{4 F \cos(\Omega t)}\right]\Bigg\rbrace,
\end{align}
where $\alpha_s^{\rm med}$ differs from its vacuum analog, and we have introduced some auxiliary variables
\begin{subequations}
\begin{align}
\kappa^2 &= \frac{k^2_\perp}{\hat q L}, \quad M^2 = \frac{\omega^2 m^2}{E^2 \hat q L}, \\
\gamma &= \frac{\hat q L^2}{2 \omega}, \quad \Omega = \frac{1-i}{\sqrt 2} \sqrt \gamma,
\end{align}
\end{subequations}
and some auxiliary functions
\begin{subequations}
\begin{align}
A &= \frac{\Omega}{4\gamma\sin[\Omega (\bar t-t)]}, \quad B = \cos[\Omega(\bar t-t)],\\
 D & = \frac{1- \bar t}{4}, \quad F = \frac{\Omega}{4\gamma \sin(\Omega t)}.
\end{align}
\end{subequations}

Unlike previous jet substructure calculations in the BDMPS-Z framework~\cite{Mehtar-Tani:2016aco,Caucal:2021cfb}, we do not take the $\omega\ll\omega_c$ and $k_t\ll Q_s$ limit that leads to a simplified, factorised formula for the spectrum but rather use the full expression. We do so in order to avoid artificial cuts in the phase-space, e.g. generating gluons with only $\omega<\omega_c$, that could alter the qualitative picture we aim at. The typical formation time of these medium-induced emissions can be estimated by setting $k^2_t=\hat q t_f$ in Eq.~\eqref{eq:tf-vac} and solving for $t_f$. We find
\begin{align}
\label{eq:tfmed-massive}
 t_f^{\rm med} &= \frac{\sqrt{8\hat q \omega + \omega^4 \theta_0^4}-\theta_0^2\omega^2}{2\hat q} \\
&\approx \sqrt{\frac{2\omega}{\hat q}} - \frac{\omega^2\theta_0^2}{2\hat q} + \frac{\theta_0^4}{8}\sqrt{\frac{\omega^7}{2\hat q^3}}+ \mathcal{O}(\theta_0^8), \nonumber
\end{align}
where in the second line we have expanded in the small-$\theta_0$ limit. Note that the first term in the expansion coincides with the massless limit and the corrections proportional to the $\theta_0$ come with alternating signs. We
observe that, similar to the vacuum case, the formation time is reduced by the mass of the emitter.

In Fig.~\ref{fig:lp-medium} we show the Lund plane density corresponding to Eq.~\eqref{eq:spectrum-massive} on the top panel and for the DLA vacuum baseline in the bottom one. Let us first focus on the purely medium induced result. The first remark is that the quantity represented in Fig.~\ref{fig:lp-medium} is unphysical since a medium-induced cascade can only develop alongside vacuum fragmentation. Thus, the fact that the spectrum becomes negative in certain regions of phase-space simply means that medium induced corrections reduce the vacuum branching probability in these zones. We observe radiation clumps around the typical transverse momentum scale acquired by diffusion in transverse space $Q^2_s=\hat q L$. In turn, the hard-collinear region, beyond the regime of validity of Eq.~\eqref{eq:spectrum-massive}, is depleted both for light and charm quarks. Regarding the DLA vacuum result, we observe that the inclusive result is purely homogeneous, while for heavy-quarks the dead cone effect clearly reduces collinear radiation.

 \begin{figure*}[t]
    \centering
    \includegraphics[width=0.99\textwidth]{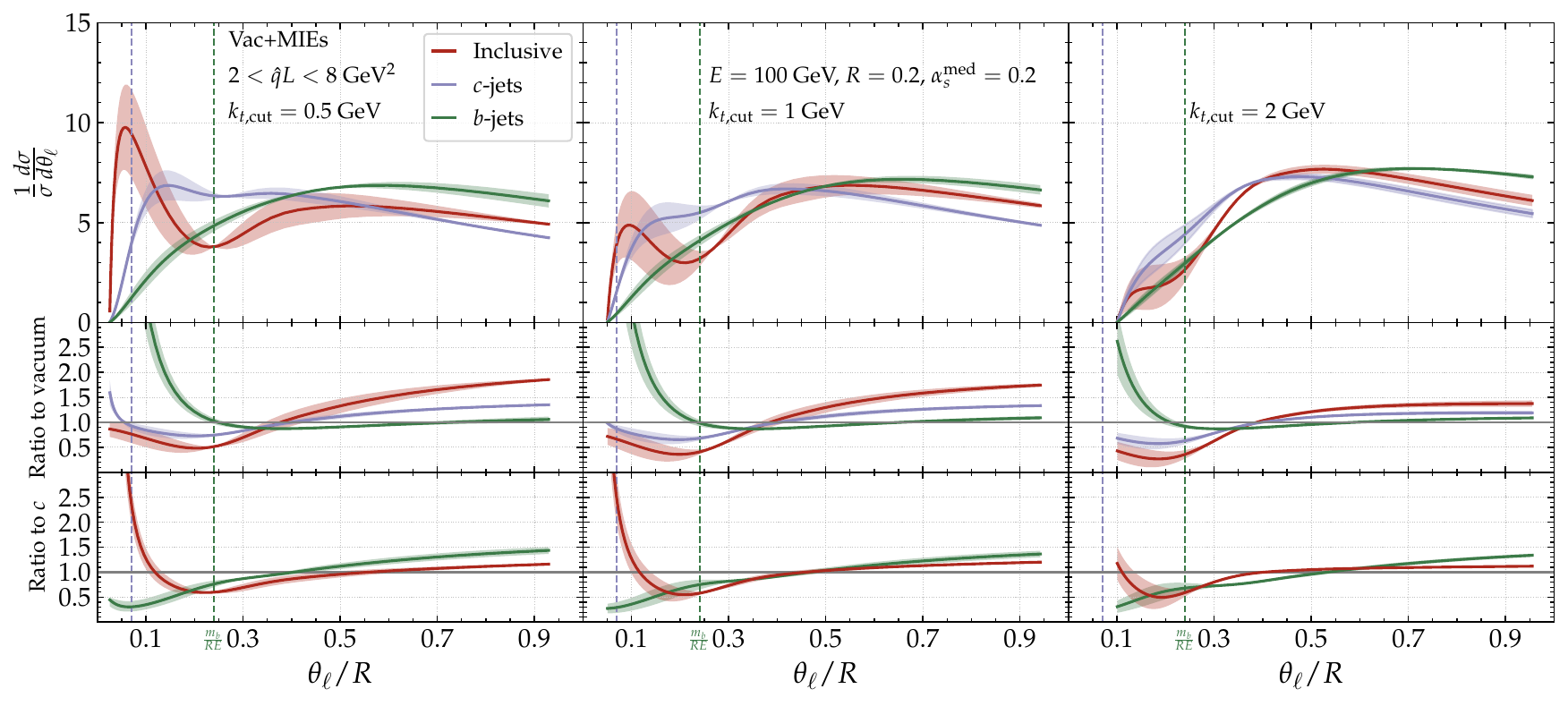}
    \caption{Top panel: same as Fig.~\ref{fig:thg-vacuum} but including both vacuum and medium induced emissions. The bottom panels display the ratio to vacuum and to in-medium charm jets. The vertical
      dashed lines denote the positions of the dead cone angles of the first splitting of the jet tree. The band is generated by varying the value of $\hat q L$ from 2 to 8 GeV$^2$.}
    \label{fig:thg-medium}
\end{figure*}

Even if no quantitative statements can be drawn from these Lund planes due to the inherent simplifications of our jet-medium interaction model, they are helpful to understand where the signal of medium-induced gluons occurs in phase-space and consequently which challenges might have to be faced in an experimental measurement. A key feature of Fig.~\ref{fig:lp-medium} is that in the case of $b$-jets, we observe medium-induced gluon radiation with angles $\theta<\theta_0$ in regions of phase-space where the vacuum distribution vanishes, while this is not the case for $c$-jets. This is a result of the interplay between the minimum angle a medium induced emission can have due to transverse momentum broadening ($\theta_c\propto 1/\sqrt{\hat q L^3}$ in the massless case), and the dead cone angle.~\footnote{We note that the value of $\theta_c$ receives mass corrections as was already noted in Ref.~\cite{Armesto:2011ir}. We leave a more through study of the mass corrections to both $\theta_c$ and the so-called veto region~\cite{Caucal:2018dla} for a separate publication.} In the case of $c$-jets, the dead cone angle is very small and, as we observe in the Lund plane, there is no effective region of phase space between the $\theta=\theta_0$ and the $z=1$ lines. Conversely, a bottom quark has a larger dead cone angle and thus a finite region of phase-space exists between $\theta_0$ and $\theta_c$ that enables medium induced emissions in the $\theta_0<\theta<\theta_c$ region. This hierarchy between $\theta_0$ and $\theta_c$ is crucial for medium induced emissions to fill the dead cone. For this concrete choice of medium parameters, the dead-cone is filled by relatively soft gluons. In fact, in agreement with Ref.~\cite{Aurenche:2009dj}, the formation time of this gluons is comparable with, or even exceeds, the size of the plasma. It would be very informative to check how does this qualitative picture changes with a more accurate calculation of the medium-induced spectrum. To that end, recent efforts, both semi-analytic~\cite{Barata:2021wuf,Isaksen:2022pkj,Blok:2020jgo} and numerical~\cite{Feal:2018sml,Andres:2020vxs,Andres:2020kfg}, ought to be extended to the massive case.

We conclude this analytic section by examining the $\theta_{\ell}$-distribution
when both vacuum and medium induced emissions are taken into account. In this case, our tagged emission can be either vacuum or medium induced and we have to veto over both types of branching at angles smaller than $\theta_{\ell}$. This is achieved by re-writing Eq.~\eqref{eq:vac-thetag} as:
\begin{equation}
\label{eq:thetag-medium}
     \frac{1}{\sigma}\frac{\dd \sigma}{\dd \theta_{\ell}}=
 \int_0^{1}\dd z
    \mathcal{P}^{\text {med}}(z,\theta_{\ell})\Theta(z\theta_{\ell} -
    \barktcut)\, \Sigma^{\text{med}}(\theta_{\ell}),
\end{equation}
with
\begin{equation}
\mathcal{P}^{\text{med}}\Sigma^{\text{med}} = (\mathcal{P}^{\text{vac}}+ \mathcal{P}^{\text{mie}}) \Sigma^{\text{vac}}\Sigma^{\text{mie}},
\end{equation}
and $\mathcal{P}^{\text {vac}}, \mathcal{P}^{\text{mie}}$ corresponding to Eq.~\eqref{eq:branch-kernel-vac} in the $z\to 0$ limit and to the medium-induced spectrum given by Eq.~\eqref{eq:spectrum-massive} (suitably transformed to ($z,\theta$) coordinates), respectively. The relation between any of the two branching kernels and their Sudakov form factor is still given by Eq.~\eqref{eq:vac-sigma-thetag}.

The results of evaluating Eq.~\eqref{eq:thetag-medium} for inclusive, $c$- and $b$-jets are shown in Fig.~\ref{fig:thg-medium}. For inclusive and charm jets, we observe an enhancement of wide-angle emissions due to transverse momentum broadening for all values of $\ktcut$. In turn, $b$-initiated jets show a  a significant excess of collinear radiation when compared to the purely vacuum curves, in agreement with the discussion around the Lund planes of Fig.~\ref{fig:lp-medium}. We have thus shown that, at least within this model, the $\theta_{\ell}$-distribution can expose the filling of the dead cone by medium induced emissions in $b$-initiated jets, while it reveals transverse momentum broadening for light and $c$-initiated jets.

 \begin{figure*}[t]
   \centering{
     \includegraphics[width=0.99\textwidth]{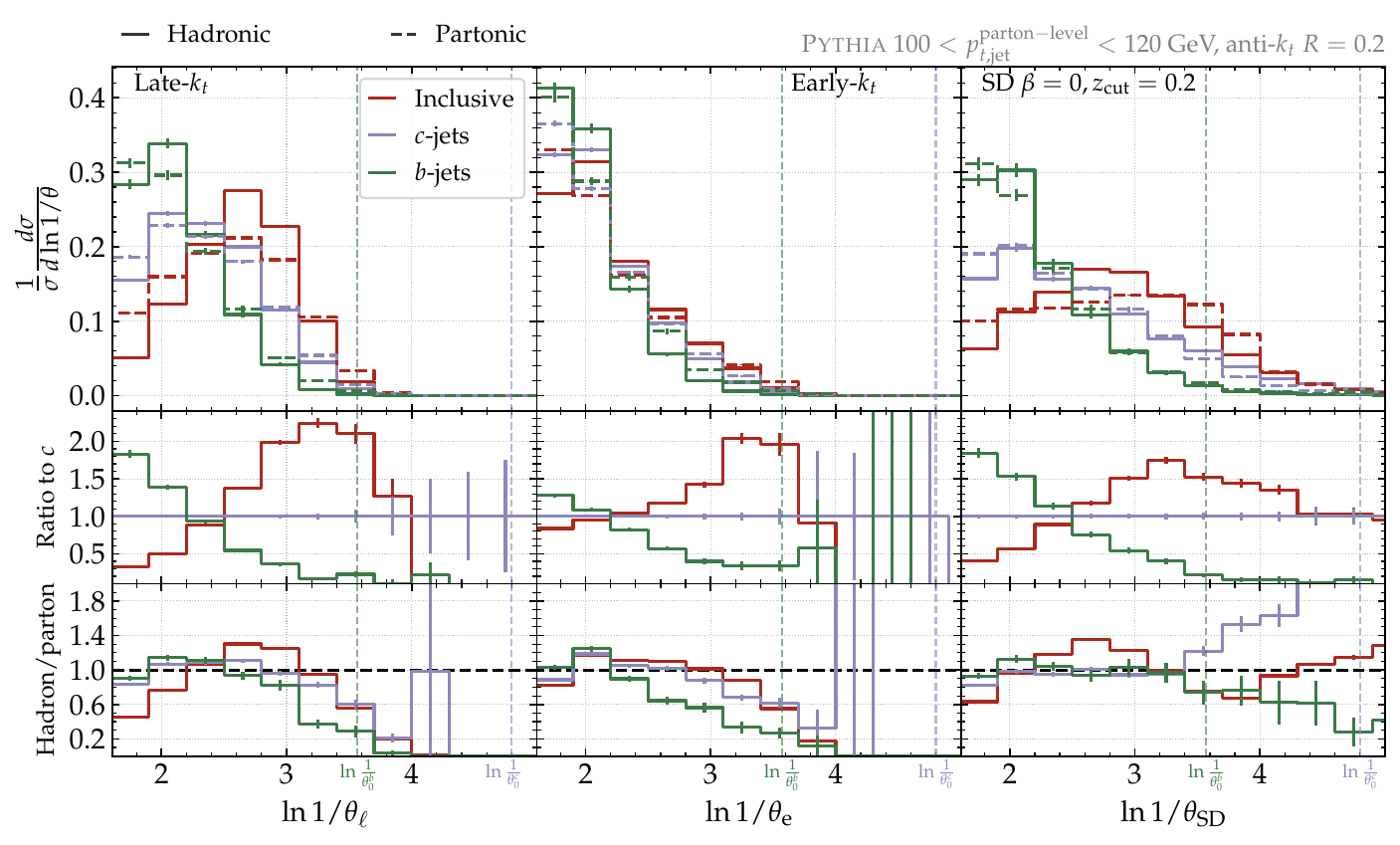}   }
    \caption{Angular distribution of the splitting tagged by \Latekt~(left), Early-$k_t$ (center) and Soft Drop (right) for
      inclusive jets (red), $c$-jets (purple) and $b$-jets (green) simulated
      with \Pythia8 at parton (dashed) and hadron (solid) level. We report in the two lower panels the
      ratio of the hadronic curves to the $c$-jet case (top) and the ratio of hadronic jets over partonic jets (bottom).}
    \label{fig:thg-vacuum-pythia}
\end{figure*}

An important aspect that we have not yet discussed is the impact of energy loss on these results. The mass-dependence of both collisional and radiative energy loss has been discussed in several works in the literature~\cite{Djordjevic:2003zk,Djordjevic:2004nq,Mustafa:2004dr,Ke:2018tsh,Herzog:2006gh,Sievert:2019cwq,Blok:2020kkn,Armesto:2005iq}. It has been shown that collisional energy loss can be safely neglected for small jet radius and moderately high-$p_t$ jets such as the ones considered in this work~\cite{Huang:2013vaa, Mustafa:2004dr}. Concerning radiative energy loss, there are two facets to consider: its color charge and its mass dependence. We propose to get rid of the former by comparing the $\theta_\ell$-distribution of $c$- and $b$-initiated jets, instead of the standard choice of taking the inclusive results as a baseline~\cite{CMS:2013qak,ATLAS:2022fgb,CMS:2022btc}. Regarding the mass dependence, the question of whether a $b$-initiated jet loses more or less energy than a $c$-initiated one is yet to be settled either theoretically or experimentally.~\footnote{An experimental measurement of the nuclear modification factor, $R_{AA}$, for $c$-tagged jets has not been put forward. Only $b$-tagged jets have been compared to inclusive jets in Refs.~\cite{ CMS:2013qak,ATLAS:2022fgb,CMS:2022btc} and, as we have already mentioned, this comparison is sensitive to both the color charge and the mass of the radiator.}  Again, the mass-dependent restriction of phase-space for medium induced emissions~\cite{Ke:2020nsm,Li:2018xuv,Huang:2013vaa} competes with the filling of the dead cone effect that we have described~\cite{Aurenche:2009dj,Armesto:2003jh}. In addition, one has to also take into account that the partonic spectrum of massive quarks is less steep than that of light quarks and, therefore, heavy quarks are less sensitive to $p_t$-migration effects. The absolute magnitude of energy loss for $c,b$-jets is an important piece of information, but what may change the qualitative picture depicted in Fig.~\ref{fig:thg-medium} is actually the dependence of energy loss on the opening angle of the tagged splitting. Since narrow, unresolved splittings ($\theta<\theta_c$) are quenched less than wide ones, all $\theta_\ell$ distributions after quenching would be narrower. This could potentially lead to the $c$-jets curve also showing an enhancement at small angles (despite being smaller than the $b$-jets one). Then, the difference between $c$ and $b$ could not uniquely be attributed to medium induced emissions filling the dead cone, but also to color decoherence dynamics. The relative weight of these two contributions could then be pined down with a centrality- (since $\theta_c \propto 1/L^3$) and energy-scan (since $\theta_0 \propto 1/E$) and it's beyond the scope of this work.

\section{Monte Carlo studies}
\label{sec:experimental}

The last part of this work is dedicated to further studying the properties of the $\theta_\ell$-distribution with realistic Monte Carlo simulations. We address three different questions: (i) whether the mass hierarchy observed in our analytic results of Fig.~\ref{fig:thg-vacuum} survives after including hadronisation effects, (ii) the impact of the fluctuating thermal background on the $\theta_\ell$ distribution and, in particular, in the dead cone regime, and (iii) make a quantitative comparison of the performance of Late-$k_t$ with respect to two other grooming strategies. On the one hand, we explore the de facto standard groomer in jet substructure studies in heavy-ions, namely Soft Drop with $\beta=0$ and $z_{\rm cut} = 0.2$. On the other hand, to emphasize the importance of selecting the most collinear splitting in the clustering sequence, we implement another groomer, that we refer to as Early-$k_t$, in which we tag the first splitting that satisfies the $\ktcut$. The simulation setup is identical to that of Sec.~\ref{sec:latekt}.

\subsection{$pp$ baseline}

\begin{figure*}[t]
   \centering{
     \includegraphics[width=0.49\textwidth]{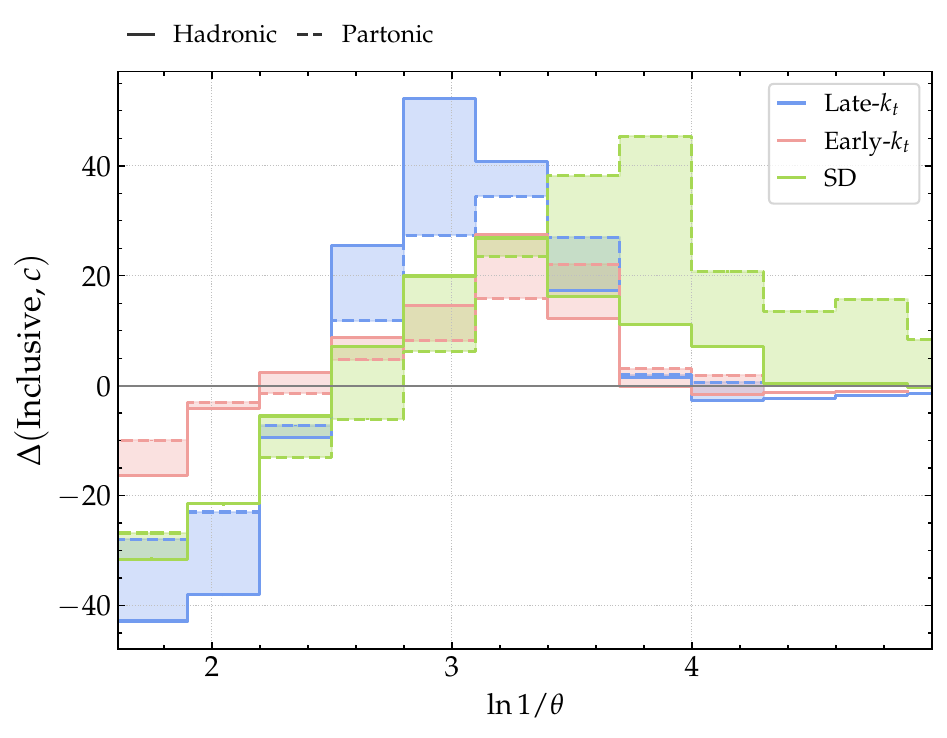}
     \includegraphics[width=0.49\textwidth]{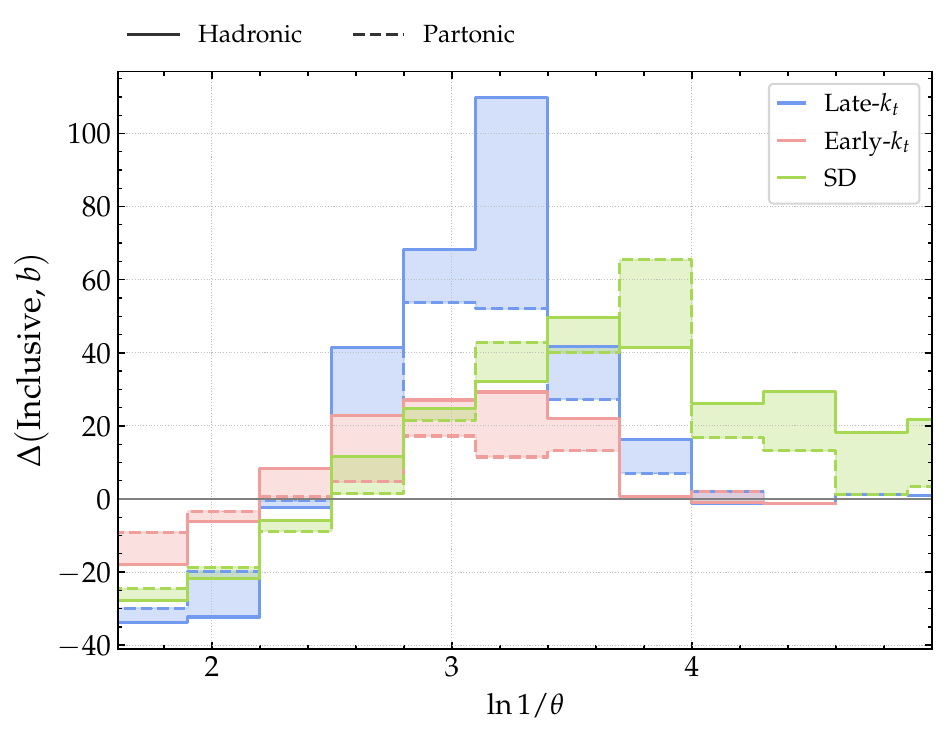}
    }
    \caption{Separation between $c$- (left), $b$-initiated (right) jets and the inclusive sample as
      defined in Eq.~(\ref{eq:separation}) for all three groomers at hadron (solid) and parton (dashed) level. The size of the colored band indicates the impact of non-perturbative corrections.}
    \label{fig:separation}
\end{figure*}

The angular distribution of the groomed splitting for each of the taggers are shown in Fig.~\ref{fig:thg-vacuum-pythia} for the different flavours and both parton and hadron level. Let us first focus on the \Latekt~distributions. Both at partonic and hadronic level, we observe qualitative agreement with the analytic study: at small $\theta_\ell$ (correspondingly large values of $\ln(1/\theta_\ell)$) the inclusive distribution clearly surpasses that of heavy flavour, as expected from dead cone effects. This is further quantified in the first ratio plot where we see that the $b$-initiated distribution is depleted by 80\% at small angles compared to the $c$-jet case. The impact of hadronisation is assessed in the second panel. The overall size of hadronisation corrections highly depends on the angular region, i.e. it varies from being less than 20\% for $0.2 <\theta_\ell/R <0.65$ to 50\% for emissions close to the jet boundary in the inclusive case. For $\theta_\ell<0.2$ we observe a strong flavour dependence of the hadronisation corrections, being the $b$-initiated jet the one that receives the largest non-perturbative correction. We conclude that hadronisation corrections to this observable, despite being too large for a pure pQCD-to-data comparison, remains sufficiently small so as to preserve the mass ordering. We would like to emphasise that for $pp$ collisions, by definition, the reach of the $\theta_{\ell}$ distribution to expose the dead cone is smaller than the observable used in the ALICE analysis based on the energy of the radiator. However, it is sufficiently good to expose dead cone dynamics in $pp$ and, in addition, it is robust to the background of a heavy-ion environment, as we show in what follows.

Concerning the comparison to the other two groomers, we first discuss the
differences between selecting the first or the last splitting. Naturally, the
Early-$k_t$ distributions peak at larger angles and consequently this leads to
very little discriminating power (below 10\%) in the large angle region, when
taking the ratio with respect to $c$-jets. Regarding hadronisation corrections,
we obtain very similar results to that of Late-$k_t$, with the biggest
difference being the 40\% reduction on the first bin for the inclusive
case. Turning to the Soft Drop results, we find that for inclusive and charm
quarks it tags emissions at smaller angles than Late-$k_t$, while the bottom curve is very similar in both cases. This is expected since the $b$-fragmentation is relatively short and thus less affected by which splitting is tagged. We observe that the discriminating power for $c$-jets is significantly reduced and that non-perturbative corrections generate an enhancement of collinear splittings both for inclusive and charm jets. This second fact is a particularly worrisome feature of this groomer since it darkens the dead cone effect.

\begin{figure*}[t]
  \centering{
    \includegraphics[width=0.99\textwidth]{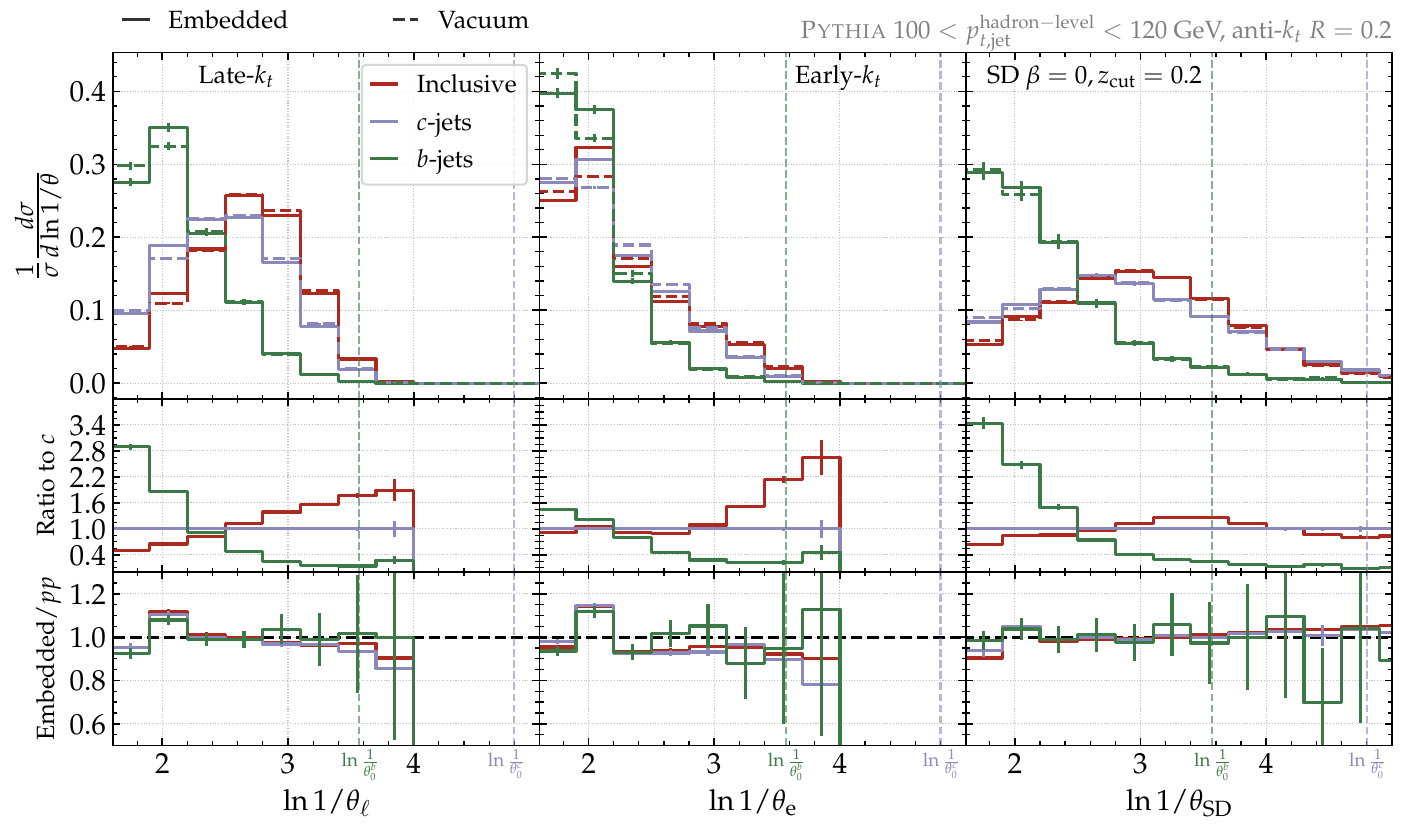}
    }
    \caption{Same as Fig.~\ref{fig:thg-vacuum-pythia} but comparing hadronic jets in $pp$ with hadronic jets created in $pp$ and embedded in a thermal background. We report in the bottom panels the ratio of the embedded curves with respect to $c$-jets (top) and quantify the impact of thermal background effects by presenting the ratio of embedded over $pp$ jets (bottom). Note that the jet-$p_t$ selection is now done at hadron level, while in Fig.~\ref{fig:thg-vacuum-pythia} it was done at parton level.}
    \label{fig:thg-thermal}
\end{figure*}

To further characterise the performance of each tagger we introduce a `separation metric'
defined as
\begin{equation}
  \label{eq:separation}
  \Delta(x,y) = \frac{x-y}{\sqrt{\sigma_x^2 + \sigma_y^2}}\,,
\end{equation}
where $x$ and $y$ are the two jet samples under study and $\sigma$ corresponds
to the statistical uncertainty. The larger the value of $\Delta(x,y)$ is, the
more discriminating power a tagger has. The result of evaluating Eq.~\eqref{eq:separation}
for the \Pythia~samples of inclusive and heavy-flavour jets is displayed in Fig.~\ref{fig:separation}.
Both for $c$ and $b$ jets, Early-$k_t$ yields the smallest value of $\Delta$ for all opening angles
and thus its performance is poor. In turn, \Latekt~results into the largest separation
for opening angles larger than $\theta \sim 0.03$ when the jet is initiated by a charm quark. We also
note that hadronisation corrections enhance the discriminating power of this groomer, while they reduce Soft
Drop's performance. The situation is very similar for $b$-initiated jets although the transition
from \Latekt~to Soft Drop being the one with largest $\Delta$ occurs at slightly larger angles
$\theta \sim 0.04$. Overall, these plots confirm the sensitivity of the new Late-$k_t$ groomer
to mass effects and that they are not washed out by non-perturbative corrections.

\subsection{Impact of thermal background}
\label{sec:thermal}

In order to test the impact of the underlying event in heavy-ion collisions on the $\theta_{\ell}$-distribution, we construct a thermal model with global properties similar to those of a realistic event. We generate randomly $N$ particles from a Gaussian distribution with average $\langle dN/d\eta\rangle =2700$, which corresponds approximately to a $0-10\%$ central collision. The $p_{t}$ of the particles is sampled from a Gamma distribution.\footnote{$\Gamma(x)=x^{\alpha-1}e^{-\beta x}$ with $\alpha=2$ and $\beta=0.35$.} This setting generates a fluctuating background with average transverse momentum per-unit-area $\langle\rho\rangle \approx 270$ GeV and a residuals distribution of the jet $p_{t}$ with width $\sigma (\delta_{p_{t}}) \approx 6.7$ GeV, similar to what is found in data~\cite{ALICE:2019qyj}. We then create hybrid events in which the \Pythia~event is mixed with the thermal background. We first run event-wise constituent subtraction on such hybrid events~\cite{Berta:2019hnj}. Then, we geometrically match the original \Pythia~jets (probe jets) to the jets reconstructed in the hybrid events (embedded jets) and study how the thermal background distorts the observable.
We further impose a minimum transverse momentum of $p_t>0.15$ GeV at particle level, to mimic realistic experimental conditions.

In Fig.~\ref{fig:thg-thermal} we show the angular distribution at vacuum and embedded level for all three groomers. In general, we find small or negligible effects due to the residual background fluctuations, after background subtraction. We note that the mass hierarchy is reduced when binning in hadron jet momentum (middle plot) as compared to binning in parton jet momentum as in Fig.~\ref{fig:thg-vacuum-pythia}. We have checked that this is just a binning artefact. 

\begin{figure*}
 \includegraphics[width=0.48\textwidth]{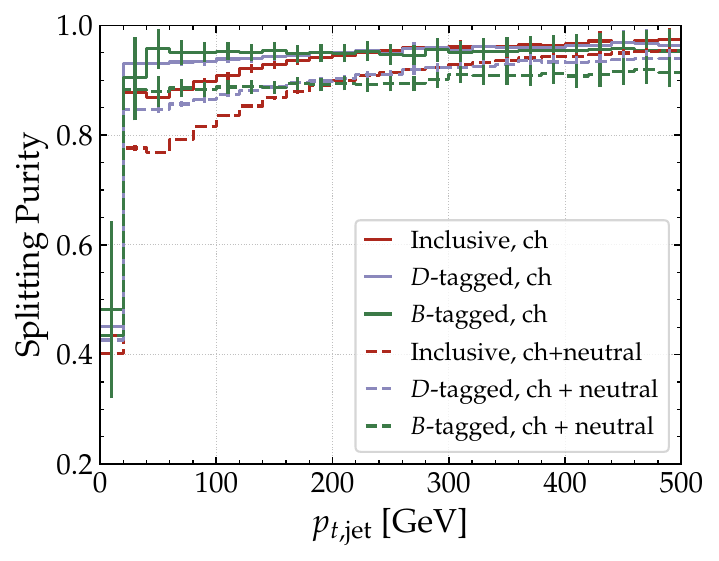}
  \includegraphics[width=0.48\textwidth]{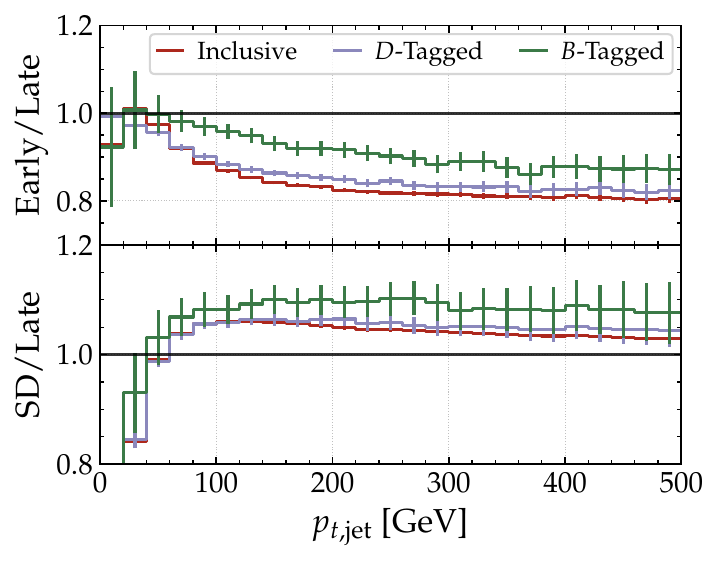}
    \caption{Left: comparison of the splitting purity in $0-10\%$ central collisions as function of the probe jet $p_{t}$ for \Latekt. Right: ratio of splitting purities between Early-$k_t$ and Late-$k_t$ (top) and the analogous result for Soft Drop (bottom) using all particles.}
    \label{fig:purity}
\end{figure*}

A robust way of estimating the distortion due to underlying event is to quantify the purity of the splittings~\cite{Mulligan:2020tim}, i.e. to check whether the two splitting prongs given by a given groomer in the embedded, matched jet correspond to the two splitting prongs in the \Pythia~probe jet. We declare correspondence when at least 50\% of the embedded prong momentum is carried by probe prong constituents. If the purity is too low -mostly due to fake subleading prongs in the embedded jet due to background-, the observable can't be corrected to particle level via unfolding since these fake splittings generate non-diagonal elements in the response matrix. More concretely, fully corrected jet substructure measurements in heavy-ion collisions have been achieved for purities larger than 80\%. The resulting purities of \Latekt~are shown in the left panel of Fig.~\ref{fig:purity} as a function of the reconstructed jet $p_{t}$ for different jet flavours. We also include the purity values corresponding to a track-based measurement scenario. Purities are above $82\%$ (and about $10\%$ higher for the track-based case) at $100$ GeV and increase towards $100\%$ as the jet $p_{t}$ increases. The main source of impurity are the cases when the leading probe prong is unevenly distributed among the embedded prongs and the cases when the whole probe splitting is contained in the embedded leading prong. These high purities guarantee that small residual differences due to background on the $\theta_{\ell}$-distribution observed in Fig.~\ref{fig:thg-thermal} would be corrected via unfolding together with detector effects in the experimental measurement. As expected from the results of Fig.~\ref{fig:thg-thermal}, the purities are lower for Early-$k_t$ and around a 5-10\% higher for Soft Drop due to the $z_{\rm cut}$.  Based on all these observations, we conclude that Late-$k_t$ is suited for dead cone searches in heavy ion collisions.

\section{Conclusions}
\label{sec:conclusions}
The long sought experimental search for the dead-cone effect in QCD has recently come to an end~\cite{ALICE:2021aqk}. The use of modern iterative declustering techniques in heavy-flavour jets~\cite{Cunqueiro:2018jbh} was instrumental in that discovery since they provide access to the energy of the radiator and, therefore, directly test Eq.~\eqref{eq:deadcone-def}. Due to the noisy environment of heavy-ion collisions, such a clean measurement cannot be replicated with the current technology and novel observables with low sensitivity to the large and fluctuating thermal background need to be devised. Potential candidates must satisfy at least three criteria: (i) strong discriminating power between light and heavy quark results, (ii) reduced sensitivity to hadronisation effects which could potentially populate the dead cone, and (iii) resilience to uncorrelated thermal background such that the observable can be corrected to particle level via unfolding.

From a theoretical viewpoint, searches of mass effects in jet observables in heavy-ion collisions are motivated by the fact that suppression of collinear radiation due to the dead cone angle can be overcome by medium induced emissions whose formation time is of the order of the Quark Gluon Plasma length~\cite{Armesto:2003jh,Aurenche:2009dj}. More concretely, multiple scatterings between a highly energetic quark and the Quark Gluon Plasma trigger additional radiation with typical energies and angles that are distinct from vacuum ones. In the massive case, these medium induced emissions can appear inside the dead cone if the minimal angle set by transverse momentum broadening, $\theta_c$, is smaller than the dead cone angle, i.e. $\theta_0 > \theta_c$. The ordering between these two scales depends on the properties of both the medium ($\hat q , L$) and the radiating parton ($m_Q, E_{\text {rad}}$). Then, an experimental measurement sensitive to the collinear regime for different jet flavours would represent a step forward in our understanding of the plasma properties and its interaction with hard propagating probes.

In this paper, we propose to use jet grooming techniques to identify a particular splitting in the jet tree that is both perturbative and sensitive to dead cone effects. We choose the most collinear splitting along the jet primary branch whose relative transverse momentum is above a certain cutoff, $\ktcut\sim\mathcal{O}(1)$ GeV. In the case of heavy-flavour jets we also impose that the leading prong contains at least one heavy-flavour hadron. Since this splitting will typically occur deep in the declustering sequence, for a low enough value of $\ktcut$, we dub this grooming technique \Latekt. Our proposed observable is the angular distribution of the splitting tagged by \Latekt~that we denote as $\theta_{\ell}$. An important result of this paper is that selecting the most collinear splitting instead of following the common practice of selecting the first, e.g. as in Soft Drop, enhances dead cone effects. In fact, while for $b$-initiated jets selecting the first or the last splitting plays a small role due to its short fragmentation, the tagged splitting choice makes a significant difference for charm and inclusive jets and increases the separation between the different flavours. At the same time, selecting the most collinear splitting reduces the impact of uncorrelated thermal background that typically manifests as fake large angle splittings in the clustering sequence. We hence conclude that the $\theta_\ell$-distribution is a robust observable from an experimental perspective.

On the theoretical side, we analyse the resummation structure of the
$\theta_{\ell}$-distribution up to modified leading logarithmic accuracy in
proton-proton collisions. A natural concern is that tagging the most collinear
splitting might lead to collinear unsafe observables in the massless
case. However, tagging the most collinear splitting is not a problem since we
also impose a $\ktcut$. The value of this free parameter has to be chosen with
particular care since it can generate an angular cutoff $\propto \ktcut/p_t$
that can mimic a dead cone for the inclusive case. We carry over this
exploratory study of the resummation structure of this observable to jets in
heavy-ions by including medium induced emissions triggered by multiple soft
scatterings with the QGP. To that end, we use the differential spectrum that was
calculated in Ref.~\cite{Armesto:2003jh}. We find that the
$\theta_\ell$-distribution for $b$-initiated jets is enhanced in the small angle
regime (below $\theta_0$) compared to the vacuum baseline, being a neat
indication of the filling of the dead cone by medium induced emissions. In turn,
the curves for inclusive and $c$-initiated jets display an enhancement at large
angles due to transverse momentum broadening. These analytic results are not
meant to be a quantitative prediction, but rather serve the purpose of
illustrating how a well-known property of the medium induced spectrum for heavy
quarks, namely the generation of gluons inside the dead cone, is imprinted in a
jet substructure observable that can eventually be measured experimentally.

This paper is just the first step towards our long term goal of bringing our understanding of jet substructure in heavy ions to a comparable level to that of proton-proton both for massless and massive quarks. A natural continuation of this work would be to resum the $\theta_\ell$-distribution to next-to-leading logarithmic accuracy. Very recently an alternative observable based on energy-energy correlators has been proposed for exposing the dead cone effect in proton-proton
collisions~\cite{Craft:2022kdo}. It would be interesting to make a systematic comparison of the pros and cons of clustering based observables with respect to energy-energy correlators, not only for dead cone searches, both in $pp$ and in heavy-ions~\cite{Andres:2022ovj}. Another important piece of information that could complement the analytic estimates presented in this work, is the result for the $\theta_\ell$-distribution obtained from jet quenching Monte Carlo codes that incorporate some subset of mass effects~\cite{Zigic:2018smz,JETSCAPE:2022hcb,Ke:2020nsm}.

\section*{ACKNOWLEDGEMENTS}
We would like to express our gratitude to Cristian Baldenegro, João Barata, Paul Caucal and Konrad
Tywoniuk for helpful discussions during the realisation of this work. A.S.O.’s
work was supported by the European Research Council (ERC) under the European
Union’s Horizon 2020 research and innovation programme (grant agreement
No. 788223, PanScales). LCM was supported by the European Research Council
project ERC-2020-COG-101002207 QCDHighDensityCMS. %
The work of DN
is supported by the ERC Starting Grant 714788 REINVENT.%

\bibliography{biblio}

\begin{thebibliography}{79}%
\makeatletter
\providecommand \@ifxundefined [1]{%
 \@ifx{#1\undefined}
}%
\providecommand \@ifnum [1]{%
 \ifnum #1\expandafter \@firstoftwo
 \else \expandafter \@secondoftwo
 \fi
}%
\providecommand \@ifx [1]{%
 \ifx #1\expandafter \@firstoftwo
 \else \expandafter \@secondoftwo
 \fi
}%
\providecommand \natexlab [1]{#1}%
\providecommand \enquote  [1]{``#1''}%
\providecommand \bibnamefont  [1]{#1}%
\providecommand \bibfnamefont [1]{#1}%
\providecommand \citenamefont [1]{#1}%
\providecommand \href@noop [0]{\@secondoftwo}%
\providecommand \href [0]{\begingroup \@sanitize@url \@href}%
\providecommand \@href[1]{\@@startlink{#1}\@@href}%
\providecommand \@@href[1]{\endgroup#1\@@endlink}%
\providecommand \@sanitize@url [0]{\catcode `\\12\catcode `\$12\catcode
  `\&12\catcode `\#12\catcode `\^12\catcode `\_12\catcode `\%12\relax}%
\providecommand \@@startlink[1]{}%
\providecommand \@@endlink[0]{}%
\providecommand \url  [0]{\begingroup\@sanitize@url \@url }%
\providecommand \@url [1]{\endgroup\@href {#1}{\urlprefix }}%
\providecommand \urlprefix  [0]{URL }%
\providecommand \Eprint [0]{\href }%
\providecommand \doibase [0]{http://dx.doi.org/}%
\providecommand \selectlanguage [0]{\@gobble}%
\providecommand \bibinfo  [0]{\@secondoftwo}%
\providecommand \bibfield  [0]{\@secondoftwo}%
\providecommand \translation [1]{[#1]}%
\providecommand \BibitemOpen [0]{}%
\providecommand \bibitemStop [0]{}%
\providecommand \bibitemNoStop [0]{.\EOS\space}%
\providecommand \EOS [0]{\spacefactor3000\relax}%
\providecommand \BibitemShut  [1]{\csname bibitem#1\endcsname}%
\let\auto@bib@innerbib\@empty
\bibitem [{\citenamefont {Marchesini}\ and\ \citenamefont
  {Webber}(1990)}]{Marchesini:1989yk}%
  \BibitemOpen
  \bibfield  {author} {\bibinfo {author} {\bibfnamefont {G.}~\bibnamefont
  {Marchesini}}\ and\ \bibinfo {author} {\bibfnamefont {B.~R.}\ \bibnamefont
  {Webber}},\ }\href {\doibase 10.1016/0550-3213(90)90310-A} {\bibfield
  {journal} {\bibinfo  {journal} {Nucl. Phys. B}\ }\textbf {\bibinfo {volume}
  {330}},\ \bibinfo {pages} {261} (\bibinfo {year} {1990})}\BibitemShut
  {NoStop}%
\bibitem [{\citenamefont {Dokshitzer}\ \emph {et~al.}(1991)\citenamefont
  {Dokshitzer}, \citenamefont {Khoze},\ and\ \citenamefont
  {Troian}}]{Dokshitzer:1991fd}%
  \BibitemOpen
  \bibfield  {author} {\bibinfo {author} {\bibfnamefont {Y.~L.}\ \bibnamefont
  {Dokshitzer}}, \bibinfo {author} {\bibfnamefont {V.~A.}\ \bibnamefont
  {Khoze}}, \ and\ \bibinfo {author} {\bibfnamefont {S.~I.}\ \bibnamefont
  {Troian}},\ }\href {\doibase 10.1088/0954-3899/17/10/023} {\bibfield
  {journal} {\bibinfo  {journal} {J. Phys. G}\ }\textbf {\bibinfo {volume}
  {17}},\ \bibinfo {pages} {1602} (\bibinfo {year} {1991})}\BibitemShut
  {NoStop}%
\bibitem [{\citenamefont {Acharya}\ \emph
  {et~al.}(2022{\natexlab{a}})\citenamefont {Acharya} \emph
  {et~al.}}]{ALICE:2021aqk}%
  \BibitemOpen
  \bibfield  {author} {\bibinfo {author} {\bibfnamefont {S.}~\bibnamefont
  {Acharya}} \emph {et~al.} (\bibinfo {collaboration} {ALICE}),\ }\href
  {\doibase 10.1038/s41586-022-04572-w} {\bibfield  {journal} {\bibinfo
  {journal} {Nature}\ }\textbf {\bibinfo {volume} {605}},\ \bibinfo {pages}
  {440} (\bibinfo {year} {2022}{\natexlab{a}})},\ \bibinfo {note} {[Erratum:
  Nature 607, E22 (2022)]},\ \Eprint {http://arxiv.org/abs/2106.05713}
  {arXiv:2106.05713 [nucl-ex]} \BibitemShut {NoStop}%
\bibitem [{\citenamefont {Cunqueiro}\ and\ \citenamefont
  {P\l{}osko\'n}(2019)}]{Cunqueiro:2018jbh}%
  \BibitemOpen
  \bibfield  {author} {\bibinfo {author} {\bibfnamefont {L.}~\bibnamefont
  {Cunqueiro}}\ and\ \bibinfo {author} {\bibfnamefont {M.}~\bibnamefont
  {P\l{}osko\'n}},\ }\href {\doibase 10.1103/PhysRevD.99.074027} {\bibfield
  {journal} {\bibinfo  {journal} {Phys. Rev. D}\ }\textbf {\bibinfo {volume}
  {99}},\ \bibinfo {pages} {074027} (\bibinfo {year} {2019})},\ \Eprint
  {http://arxiv.org/abs/1812.00102} {arXiv:1812.00102 [hep-ph]} \BibitemShut
  {NoStop}%
\bibitem [{\citenamefont {Dreyer}\ \emph {et~al.}(2018)\citenamefont {Dreyer},
  \citenamefont {Salam},\ and\ \citenamefont {Soyez}}]{Dreyer:2018nbf}%
  \BibitemOpen
  \bibfield  {author} {\bibinfo {author} {\bibfnamefont {F.~A.}\ \bibnamefont
  {Dreyer}}, \bibinfo {author} {\bibfnamefont {G.~P.}\ \bibnamefont {Salam}}, \
  and\ \bibinfo {author} {\bibfnamefont {G.}~\bibnamefont {Soyez}},\ }\href
  {\doibase 10.1007/JHEP12(2018)064} {\bibfield  {journal} {\bibinfo  {journal}
  {JHEP}\ }\textbf {\bibinfo {volume} {12}},\ \bibinfo {pages} {064} (\bibinfo
  {year} {2018})},\ \Eprint {http://arxiv.org/abs/1807.04758} {arXiv:1807.04758
  [hep-ph]} \BibitemShut {NoStop}%
\bibitem [{\citenamefont {Lifson}\ \emph {et~al.}(2020)\citenamefont {Lifson},
  \citenamefont {Salam},\ and\ \citenamefont {Soyez}}]{Lifson:2020gua}%
  \BibitemOpen
  \bibfield  {author} {\bibinfo {author} {\bibfnamefont {A.}~\bibnamefont
  {Lifson}}, \bibinfo {author} {\bibfnamefont {G.~P.}\ \bibnamefont {Salam}}, \
  and\ \bibinfo {author} {\bibfnamefont {G.}~\bibnamefont {Soyez}},\ }\href
  {\doibase 10.1007/JHEP10(2020)170} {\bibfield  {journal} {\bibinfo  {journal}
  {JHEP}\ }\textbf {\bibinfo {volume} {10}},\ \bibinfo {pages} {170} (\bibinfo
  {year} {2020})},\ \Eprint {http://arxiv.org/abs/2007.06578} {arXiv:2007.06578
  [hep-ph]} \BibitemShut {NoStop}%
\bibitem [{\citenamefont {ALICE}(2020)}]{ALICE:2020fuk}%
  \BibitemOpen
  \bibfield  {author} {\bibinfo {author} {\bibnamefont {ALICE}},\ }\href@noop
  {} {\  (\bibinfo {year} {2020})}\BibitemShut {NoStop}%
\bibitem [{\citenamefont {Aad}\ \emph {et~al.}(2020{\natexlab{a}})\citenamefont
  {Aad} \emph {et~al.}}]{ATLAS:2020bbn}%
  \BibitemOpen
  \bibfield  {author} {\bibinfo {author} {\bibfnamefont {G.}~\bibnamefont
  {Aad}} \emph {et~al.} (\bibinfo {collaboration} {ATLAS}),\ }\href {\doibase
  10.1103/PhysRevLett.124.222002} {\bibfield  {journal} {\bibinfo  {journal}
  {Phys. Rev. Lett.}\ }\textbf {\bibinfo {volume} {124}},\ \bibinfo {pages}
  {222002} (\bibinfo {year} {2020}{\natexlab{a}})},\ \Eprint
  {http://arxiv.org/abs/2004.03540} {arXiv:2004.03540 [hep-ex]} \BibitemShut
  {NoStop}%
\bibitem [{\citenamefont {Armesto}\ \emph {et~al.}(2004)\citenamefont
  {Armesto}, \citenamefont {Salgado},\ and\ \citenamefont
  {Wiedemann}}]{Armesto:2003jh}%
  \BibitemOpen
  \bibfield  {author} {\bibinfo {author} {\bibfnamefont {N.}~\bibnamefont
  {Armesto}}, \bibinfo {author} {\bibfnamefont {C.~A.}\ \bibnamefont
  {Salgado}}, \ and\ \bibinfo {author} {\bibfnamefont {U.~A.}\ \bibnamefont
  {Wiedemann}},\ }\href {\doibase 10.1103/PhysRevD.69.114003} {\bibfield
  {journal} {\bibinfo  {journal} {Phys. Rev. D}\ }\textbf {\bibinfo {volume}
  {69}},\ \bibinfo {pages} {114003} (\bibinfo {year} {2004})},\ \Eprint
  {http://arxiv.org/abs/hep-ph/0312106} {arXiv:hep-ph/0312106} \BibitemShut
  {NoStop}%
\bibitem [{\citenamefont {Dokshitzer}\ and\ \citenamefont
  {Kharzeev}(2001)}]{Dokshitzer:2001zm}%
  \BibitemOpen
  \bibfield  {author} {\bibinfo {author} {\bibfnamefont {Y.~L.}\ \bibnamefont
  {Dokshitzer}}\ and\ \bibinfo {author} {\bibfnamefont {D.~E.}\ \bibnamefont
  {Kharzeev}},\ }\href {\doibase 10.1016/S0370-2693(01)01130-3} {\bibfield
  {journal} {\bibinfo  {journal} {Phys. Lett. B}\ }\textbf {\bibinfo {volume}
  {519}},\ \bibinfo {pages} {199} (\bibinfo {year} {2001})},\ \Eprint
  {http://arxiv.org/abs/hep-ph/0106202} {arXiv:hep-ph/0106202} \BibitemShut
  {NoStop}%
\bibitem [{\citenamefont {Djordjevic}\ and\ \citenamefont
  {Gyulassy}(2004)}]{Djordjevic:2003zk}%
  \BibitemOpen
  \bibfield  {author} {\bibinfo {author} {\bibfnamefont {M.}~\bibnamefont
  {Djordjevic}}\ and\ \bibinfo {author} {\bibfnamefont {M.}~\bibnamefont
  {Gyulassy}},\ }\href {\doibase 10.1016/j.nuclphysa.2003.12.020} {\bibfield
  {journal} {\bibinfo  {journal} {Nucl. Phys. A}\ }\textbf {\bibinfo {volume}
  {733}},\ \bibinfo {pages} {265} (\bibinfo {year} {2004})},\ \Eprint
  {http://arxiv.org/abs/nucl-th/0310076} {arXiv:nucl-th/0310076} \BibitemShut
  {NoStop}%
\bibitem [{\citenamefont {Zhang}\ \emph {et~al.}(2004)\citenamefont {Zhang},
  \citenamefont {Wang},\ and\ \citenamefont {Wang}}]{Zhang:2003wk}%
  \BibitemOpen
  \bibfield  {author} {\bibinfo {author} {\bibfnamefont {B.-W.}\ \bibnamefont
  {Zhang}}, \bibinfo {author} {\bibfnamefont {E.}~\bibnamefont {Wang}}, \ and\
  \bibinfo {author} {\bibfnamefont {X.-N.}\ \bibnamefont {Wang}},\ }\href
  {\doibase 10.1103/PhysRevLett.93.072301} {\bibfield  {journal} {\bibinfo
  {journal} {Phys. Rev. Lett.}\ }\textbf {\bibinfo {volume} {93}},\ \bibinfo
  {pages} {072301} (\bibinfo {year} {2004})},\ \Eprint
  {http://arxiv.org/abs/nucl-th/0309040} {arXiv:nucl-th/0309040} \BibitemShut
  {NoStop}%
\bibitem [{\citenamefont {Aurenche}\ and\ \citenamefont
  {Zakharov}(2009)}]{Aurenche:2009dj}%
  \BibitemOpen
  \bibfield  {author} {\bibinfo {author} {\bibfnamefont {P.}~\bibnamefont
  {Aurenche}}\ and\ \bibinfo {author} {\bibfnamefont {B.~G.}\ \bibnamefont
  {Zakharov}},\ }\href {\doibase 10.1134/S0021364009160048} {\bibfield
  {journal} {\bibinfo  {journal} {JETP Lett.}\ }\textbf {\bibinfo {volume}
  {90}},\ \bibinfo {pages} {237} (\bibinfo {year} {2009})},\ \Eprint
  {http://arxiv.org/abs/0907.1918} {arXiv:0907.1918 [hep-ph]} \BibitemShut
  {NoStop}%
\bibitem [{\citenamefont {Mehtar-Tani}\ \emph {et~al.}(2011)\citenamefont
  {Mehtar-Tani}, \citenamefont {Salgado},\ and\ \citenamefont
  {Tywoniuk}}]{Mehtar-Tani:2010ebp}%
  \BibitemOpen
  \bibfield  {author} {\bibinfo {author} {\bibfnamefont {Y.}~\bibnamefont
  {Mehtar-Tani}}, \bibinfo {author} {\bibfnamefont {C.~A.}\ \bibnamefont
  {Salgado}}, \ and\ \bibinfo {author} {\bibfnamefont {K.}~\bibnamefont
  {Tywoniuk}},\ }\href {\doibase 10.1103/PhysRevLett.106.122002} {\bibfield
  {journal} {\bibinfo  {journal} {Phys. Rev. Lett.}\ }\textbf {\bibinfo
  {volume} {106}},\ \bibinfo {pages} {122002} (\bibinfo {year} {2011})},\
  \Eprint {http://arxiv.org/abs/1009.2965} {arXiv:1009.2965 [hep-ph]}
  \BibitemShut {NoStop}%
\bibitem [{\citenamefont {Mehtar-Tani}\ \emph {et~al.}(2012)\citenamefont
  {Mehtar-Tani}, \citenamefont {Salgado},\ and\ \citenamefont
  {Tywoniuk}}]{Mehtar-Tani:2011hma}%
  \BibitemOpen
  \bibfield  {author} {\bibinfo {author} {\bibfnamefont {Y.}~\bibnamefont
  {Mehtar-Tani}}, \bibinfo {author} {\bibfnamefont {C.~A.}\ \bibnamefont
  {Salgado}}, \ and\ \bibinfo {author} {\bibfnamefont {K.}~\bibnamefont
  {Tywoniuk}},\ }\href {\doibase 10.1016/j.physletb.2011.12.042} {\bibfield
  {journal} {\bibinfo  {journal} {Phys. Lett. B}\ }\textbf {\bibinfo {volume}
  {707}},\ \bibinfo {pages} {156} (\bibinfo {year} {2012})},\ \Eprint
  {http://arxiv.org/abs/1102.4317} {arXiv:1102.4317 [hep-ph]} \BibitemShut
  {NoStop}%
\bibitem [{\citenamefont {Casalderrey-Solana}\ and\ \citenamefont
  {Iancu}(2011)}]{Casalderrey-Solana:2011ule}%
  \BibitemOpen
  \bibfield  {author} {\bibinfo {author} {\bibfnamefont {J.}~\bibnamefont
  {Casalderrey-Solana}}\ and\ \bibinfo {author} {\bibfnamefont
  {E.}~\bibnamefont {Iancu}},\ }\href {\doibase 10.1007/JHEP08(2011)015}
  {\bibfield  {journal} {\bibinfo  {journal} {JHEP}\ }\textbf {\bibinfo
  {volume} {08}},\ \bibinfo {pages} {015} (\bibinfo {year} {2011})},\ \Eprint
  {http://arxiv.org/abs/1105.1760} {arXiv:1105.1760 [hep-ph]} \BibitemShut
  {NoStop}%
\bibitem [{\citenamefont {Dasgupta}\ \emph {et~al.}(2013)\citenamefont
  {Dasgupta}, \citenamefont {Fregoso}, \citenamefont {Marzani},\ and\
  \citenamefont {Salam}}]{Dasgupta:2013ihk}%
  \BibitemOpen
  \bibfield  {author} {\bibinfo {author} {\bibfnamefont {M.}~\bibnamefont
  {Dasgupta}}, \bibinfo {author} {\bibfnamefont {A.}~\bibnamefont {Fregoso}},
  \bibinfo {author} {\bibfnamefont {S.}~\bibnamefont {Marzani}}, \ and\
  \bibinfo {author} {\bibfnamefont {G.~P.}\ \bibnamefont {Salam}},\ }\href
  {\doibase 10.1007/JHEP09(2013)029} {\bibfield  {journal} {\bibinfo  {journal}
  {JHEP}\ }\textbf {\bibinfo {volume} {09}},\ \bibinfo {pages} {029} (\bibinfo
  {year} {2013})},\ \Eprint {http://arxiv.org/abs/1307.0007} {arXiv:1307.0007
  [hep-ph]} \BibitemShut {NoStop}%
\bibitem [{\citenamefont {Larkoski}\ \emph {et~al.}(2014)\citenamefont
  {Larkoski}, \citenamefont {Marzani}, \citenamefont {Soyez},\ and\
  \citenamefont {Thaler}}]{Larkoski:2014wba}%
  \BibitemOpen
  \bibfield  {author} {\bibinfo {author} {\bibfnamefont {A.~J.}\ \bibnamefont
  {Larkoski}}, \bibinfo {author} {\bibfnamefont {S.}~\bibnamefont {Marzani}},
  \bibinfo {author} {\bibfnamefont {G.}~\bibnamefont {Soyez}}, \ and\ \bibinfo
  {author} {\bibfnamefont {J.}~\bibnamefont {Thaler}},\ }\href {\doibase
  10.1007/JHEP05(2014)146} {\bibfield  {journal} {\bibinfo  {journal} {JHEP}\
  }\textbf {\bibinfo {volume} {05}},\ \bibinfo {pages} {146} (\bibinfo {year}
  {2014})},\ \Eprint {http://arxiv.org/abs/1402.2657} {arXiv:1402.2657
  [hep-ph]} \BibitemShut {NoStop}%
\bibitem [{\citenamefont {Mulligan}\ and\ \citenamefont
  {Ploskon}(2020)}]{Mulligan:2020tim}%
  \BibitemOpen
  \bibfield  {author} {\bibinfo {author} {\bibfnamefont {J.}~\bibnamefont
  {Mulligan}}\ and\ \bibinfo {author} {\bibfnamefont {M.}~\bibnamefont
  {Ploskon}},\ }\href {\doibase 10.1103/PhysRevC.102.044913} {\bibfield
  {journal} {\bibinfo  {journal} {Phys. Rev. C}\ }\textbf {\bibinfo {volume}
  {102}},\ \bibinfo {pages} {044913} (\bibinfo {year} {2020})},\ \Eprint
  {http://arxiv.org/abs/2006.01812} {arXiv:2006.01812 [hep-ph]} \BibitemShut
  {NoStop}%
\bibitem [{\citenamefont {Acharya}\ \emph
  {et~al.}(2022{\natexlab{b}})\citenamefont {Acharya} \emph
  {et~al.}}]{ALargeIonColliderExperiment:2021mqf}%
  \BibitemOpen
  \bibfield  {author} {\bibinfo {author} {\bibfnamefont {S.}~\bibnamefont
  {Acharya}} \emph {et~al.} (\bibinfo {collaboration} {A Large Ion Collider
  Experiment, ALICE}),\ }\href {\doibase 10.1103/PhysRevLett.128.102001}
  {\bibfield  {journal} {\bibinfo  {journal} {Phys. Rev. Lett.}\ }\textbf
  {\bibinfo {volume} {128}},\ \bibinfo {pages} {102001} (\bibinfo {year}
  {2022}{\natexlab{b}})},\ \Eprint {http://arxiv.org/abs/2107.12984}
  {arXiv:2107.12984 [nucl-ex]} \BibitemShut {NoStop}%
\bibitem [{\citenamefont {Marzani}\ \emph {et~al.}(2019)\citenamefont
  {Marzani}, \citenamefont {Soyez},\ and\ \citenamefont
  {Spannowsky}}]{Marzani:2019hun}%
  \BibitemOpen
  \bibfield  {author} {\bibinfo {author} {\bibfnamefont {S.}~\bibnamefont
  {Marzani}}, \bibinfo {author} {\bibfnamefont {G.}~\bibnamefont {Soyez}}, \
  and\ \bibinfo {author} {\bibfnamefont {M.}~\bibnamefont {Spannowsky}},\
  }\href {\doibase 10.1007/978-3-030-15709-8} {\emph {\bibinfo {title}
  {{Looking inside jets: an introduction to jet substructure and boosted-object
  phenomenology}}}},\ Vol.\ \bibinfo {volume} {958}\ (\bibinfo  {publisher}
  {Springer},\ \bibinfo {year} {2019})\ \Eprint
  {http://arxiv.org/abs/1901.10342} {arXiv:1901.10342 [hep-ph]} \BibitemShut
  {NoStop}%
\bibitem [{\citenamefont {Larkoski}\ \emph {et~al.}(2020)\citenamefont
  {Larkoski}, \citenamefont {Moult},\ and\ \citenamefont
  {Nachman}}]{Larkoski:2017jix}%
  \BibitemOpen
  \bibfield  {author} {\bibinfo {author} {\bibfnamefont {A.~J.}\ \bibnamefont
  {Larkoski}}, \bibinfo {author} {\bibfnamefont {I.}~\bibnamefont {Moult}}, \
  and\ \bibinfo {author} {\bibfnamefont {B.}~\bibnamefont {Nachman}},\ }\href
  {\doibase 10.1016/j.physrep.2019.11.001} {\bibfield  {journal} {\bibinfo
  {journal} {Phys. Rept.}\ }\textbf {\bibinfo {volume} {841}},\ \bibinfo
  {pages} {1} (\bibinfo {year} {2020})},\ \Eprint
  {http://arxiv.org/abs/1709.04464} {arXiv:1709.04464 [hep-ph]} \BibitemShut
  {NoStop}%
\bibitem [{\citenamefont {Cunqueiro}\ and\ \citenamefont
  {Sickles}(2022)}]{Cunqueiro:2021wls}%
  \BibitemOpen
  \bibfield  {author} {\bibinfo {author} {\bibfnamefont {L.}~\bibnamefont
  {Cunqueiro}}\ and\ \bibinfo {author} {\bibfnamefont {A.~M.}\ \bibnamefont
  {Sickles}},\ }\href {\doibase 10.1016/j.ppnp.2022.103940} {\bibfield
  {journal} {\bibinfo  {journal} {Prog. Part. Nucl. Phys.}\ }\textbf {\bibinfo
  {volume} {124}},\ \bibinfo {pages} {103940} (\bibinfo {year} {2022})},\
  \Eprint {http://arxiv.org/abs/2110.14490} {arXiv:2110.14490 [nucl-ex]}
  \BibitemShut {NoStop}%
\bibitem [{\citenamefont {Aaboud}\ \emph {et~al.}(2018)\citenamefont {Aaboud}
  \emph {et~al.}}]{ATLAS:2017zda}%
  \BibitemOpen
  \bibfield  {author} {\bibinfo {author} {\bibfnamefont {M.}~\bibnamefont
  {Aaboud}} \emph {et~al.} (\bibinfo {collaboration} {ATLAS}),\ }\href
  {\doibase 10.1103/PhysRevLett.121.092001} {\bibfield  {journal} {\bibinfo
  {journal} {Phys. Rev. Lett.}\ }\textbf {\bibinfo {volume} {121}},\ \bibinfo
  {pages} {092001} (\bibinfo {year} {2018})},\ \Eprint
  {http://arxiv.org/abs/1711.08341} {arXiv:1711.08341 [hep-ex]} \BibitemShut
  {NoStop}%
\bibitem [{\citenamefont {Aad}\ \emph {et~al.}(2020{\natexlab{b}})\citenamefont
  {Aad} \emph {et~al.}}]{ATLAS:2019mgf}%
  \BibitemOpen
  \bibfield  {author} {\bibinfo {author} {\bibfnamefont {G.}~\bibnamefont
  {Aad}} \emph {et~al.} (\bibinfo {collaboration} {ATLAS}),\ }\href {\doibase
  10.1103/PhysRevD.101.052007} {\bibfield  {journal} {\bibinfo  {journal}
  {Phys. Rev. D}\ }\textbf {\bibinfo {volume} {101}},\ \bibinfo {pages}
  {052007} (\bibinfo {year} {2020}{\natexlab{b}})},\ \Eprint
  {http://arxiv.org/abs/1912.09837} {arXiv:1912.09837 [hep-ex]} \BibitemShut
  {NoStop}%
\bibitem [{\citenamefont {Tripathee}\ \emph {et~al.}(2017)\citenamefont
  {Tripathee}, \citenamefont {Xue}, \citenamefont {Larkoski}, \citenamefont
  {Marzani},\ and\ \citenamefont {Thaler}}]{Tripathee:2017ybi}%
  \BibitemOpen
  \bibfield  {author} {\bibinfo {author} {\bibfnamefont {A.}~\bibnamefont
  {Tripathee}}, \bibinfo {author} {\bibfnamefont {W.}~\bibnamefont {Xue}},
  \bibinfo {author} {\bibfnamefont {A.}~\bibnamefont {Larkoski}}, \bibinfo
  {author} {\bibfnamefont {S.}~\bibnamefont {Marzani}}, \ and\ \bibinfo
  {author} {\bibfnamefont {J.}~\bibnamefont {Thaler}},\ }\href {\doibase
  10.1103/PhysRevD.96.074003} {\bibfield  {journal} {\bibinfo  {journal} {Phys.
  Rev. D}\ }\textbf {\bibinfo {volume} {96}},\ \bibinfo {pages} {074003}
  (\bibinfo {year} {2017})},\ \Eprint {http://arxiv.org/abs/1704.05842}
  {arXiv:1704.05842 [hep-ph]} \BibitemShut {NoStop}%
\bibitem [{\citenamefont {Adam}\ \emph {et~al.}(2020)\citenamefont {Adam} \emph
  {et~al.}}]{STAR:2020ejj}%
  \BibitemOpen
  \bibfield  {author} {\bibinfo {author} {\bibfnamefont {J.}~\bibnamefont
  {Adam}} \emph {et~al.} (\bibinfo {collaboration} {STAR}),\ }\href {\doibase
  10.1016/j.physletb.2020.135846} {\bibfield  {journal} {\bibinfo  {journal}
  {Phys. Lett. B}\ }\textbf {\bibinfo {volume} {811}},\ \bibinfo {pages}
  {135846} (\bibinfo {year} {2020})},\ \Eprint
  {http://arxiv.org/abs/2003.02114} {arXiv:2003.02114 [hep-ex]} \BibitemShut
  {NoStop}%
\bibitem [{ALI(2022)}]{ALICE:2022hyz}%
  \BibitemOpen
  \href@noop {} {\  (\bibinfo {year} {2022})},\ \Eprint
  {http://arxiv.org/abs/2204.10246} {arXiv:2204.10246 [nucl-ex]} \BibitemShut
  {NoStop}%
\bibitem [{\citenamefont {Marzani}\ \emph {et~al.}(2018)\citenamefont
  {Marzani}, \citenamefont {Schunk},\ and\ \citenamefont
  {Soyez}}]{Marzani:2017kqd}%
  \BibitemOpen
  \bibfield  {author} {\bibinfo {author} {\bibfnamefont {S.}~\bibnamefont
  {Marzani}}, \bibinfo {author} {\bibfnamefont {L.}~\bibnamefont {Schunk}}, \
  and\ \bibinfo {author} {\bibfnamefont {G.}~\bibnamefont {Soyez}},\ }\href
  {\doibase 10.1140/epjc/s10052-018-5579-5} {\bibfield  {journal} {\bibinfo
  {journal} {Eur. Phys. J. C}\ }\textbf {\bibinfo {volume} {78}},\ \bibinfo
  {pages} {96} (\bibinfo {year} {2018})},\ \Eprint
  {http://arxiv.org/abs/1712.05105} {arXiv:1712.05105 [hep-ph]} \BibitemShut
  {NoStop}%
\bibitem [{\citenamefont {Kang}\ \emph {et~al.}(2020)\citenamefont {Kang},
  \citenamefont {Lee}, \citenamefont {Liu}, \citenamefont {Neill},\ and\
  \citenamefont {Ringer}}]{Kang:2019prh}%
  \BibitemOpen
  \bibfield  {author} {\bibinfo {author} {\bibfnamefont {Z.-B.}\ \bibnamefont
  {Kang}}, \bibinfo {author} {\bibfnamefont {K.}~\bibnamefont {Lee}}, \bibinfo
  {author} {\bibfnamefont {X.}~\bibnamefont {Liu}}, \bibinfo {author}
  {\bibfnamefont {D.}~\bibnamefont {Neill}}, \ and\ \bibinfo {author}
  {\bibfnamefont {F.}~\bibnamefont {Ringer}},\ }\href {\doibase
  10.1007/JHEP02(2020)054} {\bibfield  {journal} {\bibinfo  {journal} {JHEP}\
  }\textbf {\bibinfo {volume} {02}},\ \bibinfo {pages} {054} (\bibinfo {year}
  {2020})},\ \Eprint {http://arxiv.org/abs/1908.01783} {arXiv:1908.01783
  [hep-ph]} \BibitemShut {NoStop}%
\bibitem [{\citenamefont {Cal}\ \emph {et~al.}(2022)\citenamefont {Cal},
  \citenamefont {Lee}, \citenamefont {Ringer},\ and\ \citenamefont
  {Waalewijn}}]{Cal:2021fla}%
  \BibitemOpen
  \bibfield  {author} {\bibinfo {author} {\bibfnamefont {P.}~\bibnamefont
  {Cal}}, \bibinfo {author} {\bibfnamefont {K.}~\bibnamefont {Lee}}, \bibinfo
  {author} {\bibfnamefont {F.}~\bibnamefont {Ringer}}, \ and\ \bibinfo {author}
  {\bibfnamefont {W.~J.}\ \bibnamefont {Waalewijn}},\ }\href {\doibase
  10.1016/j.physletb.2022.137390} {\bibfield  {journal} {\bibinfo  {journal}
  {Phys. Lett. B}\ }\textbf {\bibinfo {volume} {833}},\ \bibinfo {pages}
  {137390} (\bibinfo {year} {2022})},\ \Eprint
  {http://arxiv.org/abs/2106.04589} {arXiv:2106.04589 [hep-ph]} \BibitemShut
  {NoStop}%
\bibitem [{\citenamefont {Frye}\ \emph {et~al.}(2016)\citenamefont {Frye},
  \citenamefont {Larkoski}, \citenamefont {Schwartz},\ and\ \citenamefont
  {Yan}}]{Frye:2016aiz}%
  \BibitemOpen
  \bibfield  {author} {\bibinfo {author} {\bibfnamefont {C.}~\bibnamefont
  {Frye}}, \bibinfo {author} {\bibfnamefont {A.~J.}\ \bibnamefont {Larkoski}},
  \bibinfo {author} {\bibfnamefont {M.~D.}\ \bibnamefont {Schwartz}}, \ and\
  \bibinfo {author} {\bibfnamefont {K.}~\bibnamefont {Yan}},\ }\href {\doibase
  10.1007/JHEP07(2016)064} {\bibfield  {journal} {\bibinfo  {journal} {JHEP}\
  }\textbf {\bibinfo {volume} {07}},\ \bibinfo {pages} {064} (\bibinfo {year}
  {2016})},\ \Eprint {http://arxiv.org/abs/1603.09338} {arXiv:1603.09338
  [hep-ph]} \BibitemShut {NoStop}%
\bibitem [{\citenamefont {Kang}\ \emph {et~al.}(2019)\citenamefont {Kang},
  \citenamefont {Lee}, \citenamefont {Liu},\ and\ \citenamefont
  {Ringer}}]{Kang:2018vgn}%
  \BibitemOpen
  \bibfield  {author} {\bibinfo {author} {\bibfnamefont {Z.-B.}\ \bibnamefont
  {Kang}}, \bibinfo {author} {\bibfnamefont {K.}~\bibnamefont {Lee}}, \bibinfo
  {author} {\bibfnamefont {X.}~\bibnamefont {Liu}}, \ and\ \bibinfo {author}
  {\bibfnamefont {F.}~\bibnamefont {Ringer}},\ }\href {\doibase
  10.1016/j.physletb.2019.04.018} {\bibfield  {journal} {\bibinfo  {journal}
  {Phys. Lett. B}\ }\textbf {\bibinfo {volume} {793}},\ \bibinfo {pages} {41}
  (\bibinfo {year} {2019})},\ \Eprint {http://arxiv.org/abs/1811.06983}
  {arXiv:1811.06983 [hep-ph]} \BibitemShut {NoStop}%
\bibitem [{\citenamefont {Makris}\ and\ \citenamefont
  {Vaidya}(2018)}]{Makris:2018npl}%
  \BibitemOpen
  \bibfield  {author} {\bibinfo {author} {\bibfnamefont {Y.}~\bibnamefont
  {Makris}}\ and\ \bibinfo {author} {\bibfnamefont {V.}~\bibnamefont
  {Vaidya}},\ }\href {\doibase 10.1007/JHEP10(2018)019} {\bibfield  {journal}
  {\bibinfo  {journal} {JHEP}\ }\textbf {\bibinfo {volume} {10}},\ \bibinfo
  {pages} {019} (\bibinfo {year} {2018})},\ \Eprint
  {http://arxiv.org/abs/1807.09805} {arXiv:1807.09805 [hep-ph]} \BibitemShut
  {NoStop}%
\bibitem [{\citenamefont {Lee}\ \emph {et~al.}(2019)\citenamefont {Lee},
  \citenamefont {Shrivastava},\ and\ \citenamefont {Vaidya}}]{Lee:2019lge}%
  \BibitemOpen
  \bibfield  {author} {\bibinfo {author} {\bibfnamefont {C.}~\bibnamefont
  {Lee}}, \bibinfo {author} {\bibfnamefont {P.}~\bibnamefont {Shrivastava}}, \
  and\ \bibinfo {author} {\bibfnamefont {V.}~\bibnamefont {Vaidya}},\ }\href
  {\doibase 10.1007/JHEP09(2019)045} {\bibfield  {journal} {\bibinfo  {journal}
  {JHEP}\ }\textbf {\bibinfo {volume} {09}},\ \bibinfo {pages} {045} (\bibinfo
  {year} {2019})},\ \Eprint {http://arxiv.org/abs/1901.09095} {arXiv:1901.09095
  [hep-ph]} \BibitemShut {NoStop}%
\bibitem [{\citenamefont {Craft}\ \emph {et~al.}(2022)\citenamefont {Craft},
  \citenamefont {Lee}, \citenamefont {Me\c{c}aj},\ and\ \citenamefont
  {Moult}}]{Craft:2022kdo}%
  \BibitemOpen
  \bibfield  {author} {\bibinfo {author} {\bibfnamefont {E.}~\bibnamefont
  {Craft}}, \bibinfo {author} {\bibfnamefont {K.}~\bibnamefont {Lee}}, \bibinfo
  {author} {\bibfnamefont {B.}~\bibnamefont {Me\c{c}aj}}, \ and\ \bibinfo
  {author} {\bibfnamefont {I.}~\bibnamefont {Moult}},\ }\href@noop {} {\
  (\bibinfo {year} {2022})},\ \Eprint {http://arxiv.org/abs/2210.09311}
  {arXiv:2210.09311 [hep-ph]} \BibitemShut {NoStop}%
\bibitem [{\citenamefont {Li}\ and\ \citenamefont
  {Vitev}(2019{\natexlab{a}})}]{Li:2017wwc}%
  \BibitemOpen
  \bibfield  {author} {\bibinfo {author} {\bibfnamefont {H.~T.}\ \bibnamefont
  {Li}}\ and\ \bibinfo {author} {\bibfnamefont {I.}~\bibnamefont {Vitev}},\
  }\href {\doibase 10.1016/j.physletb.2019.04.052} {\bibfield  {journal}
  {\bibinfo  {journal} {Phys. Lett. B}\ }\textbf {\bibinfo {volume} {793}},\
  \bibinfo {pages} {259} (\bibinfo {year} {2019}{\natexlab{a}})},\ \Eprint
  {http://arxiv.org/abs/1801.00008} {arXiv:1801.00008 [hep-ph]} \BibitemShut
  {NoStop}%
\bibitem [{\citenamefont {Maltoni}\ \emph {et~al.}(2016)\citenamefont
  {Maltoni}, \citenamefont {Selvaggi},\ and\ \citenamefont
  {Thaler}}]{Maltoni:2016ays}%
  \BibitemOpen
  \bibfield  {author} {\bibinfo {author} {\bibfnamefont {F.}~\bibnamefont
  {Maltoni}}, \bibinfo {author} {\bibfnamefont {M.}~\bibnamefont {Selvaggi}}, \
  and\ \bibinfo {author} {\bibfnamefont {J.}~\bibnamefont {Thaler}},\ }\href
  {\doibase 10.1103/PhysRevD.94.054015} {\bibfield  {journal} {\bibinfo
  {journal} {Phys. Rev. D}\ }\textbf {\bibinfo {volume} {94}},\ \bibinfo
  {pages} {054015} (\bibinfo {year} {2016})},\ \Eprint
  {http://arxiv.org/abs/1606.03449} {arXiv:1606.03449 [hep-ph]} \BibitemShut
  {NoStop}%
\bibitem [{\citenamefont {Dokshitzer}\ \emph {et~al.}(1997)\citenamefont
  {Dokshitzer}, \citenamefont {Leder}, \citenamefont {Moretti},\ and\
  \citenamefont {Webber}}]{Dokshitzer:1997in}%
  \BibitemOpen
  \bibfield  {author} {\bibinfo {author} {\bibfnamefont {Y.~L.}\ \bibnamefont
  {Dokshitzer}}, \bibinfo {author} {\bibfnamefont {G.~D.}\ \bibnamefont
  {Leder}}, \bibinfo {author} {\bibfnamefont {S.}~\bibnamefont {Moretti}}, \
  and\ \bibinfo {author} {\bibfnamefont {B.~R.}\ \bibnamefont {Webber}},\
  }\href {\doibase 10.1088/1126-6708/1997/08/001} {\bibfield  {journal}
  {\bibinfo  {journal} {JHEP}\ }\textbf {\bibinfo {volume} {08}},\ \bibinfo
  {pages} {001} (\bibinfo {year} {1997})},\ \Eprint
  {http://arxiv.org/abs/hep-ph/9707323} {arXiv:hep-ph/9707323} \BibitemShut
  {NoStop}%
\bibitem [{\citenamefont {Wobisch}\ and\ \citenamefont
  {Wengler}(1998)}]{Wobisch:1998wt}%
  \BibitemOpen
  \bibfield  {author} {\bibinfo {author} {\bibfnamefont {M.}~\bibnamefont
  {Wobisch}}\ and\ \bibinfo {author} {\bibfnamefont {T.}~\bibnamefont
  {Wengler}},\ }in\ \href@noop {} {\emph {\bibinfo {booktitle} {{Workshop on
  Monte Carlo Generators for HERA Physics (Plenary Starting Meeting)}}}}\
  (\bibinfo {year} {1998})\ pp.\ \bibinfo {pages} {270--279},\ \Eprint
  {http://arxiv.org/abs/hep-ph/9907280} {arXiv:hep-ph/9907280} \BibitemShut
  {NoStop}%
\bibitem [{\citenamefont {Bierlich}\ \emph {et~al.}(2022)\citenamefont
  {Bierlich} \emph {et~al.}}]{Bierlich:2022pfr}%
  \BibitemOpen
  \bibfield  {author} {\bibinfo {author} {\bibfnamefont {C.}~\bibnamefont
  {Bierlich}} \emph {et~al.},\ }\href@noop {} {\  (\bibinfo {year} {2022})},\
  \Eprint {http://arxiv.org/abs/2203.11601} {arXiv:2203.11601 [hep-ph]}
  \BibitemShut {NoStop}%
\bibitem [{\citenamefont {Cacciari}\ \emph {et~al.}(2008)\citenamefont
  {Cacciari}, \citenamefont {Salam},\ and\ \citenamefont
  {Soyez}}]{Cacciari:2008gp}%
  \BibitemOpen
  \bibfield  {author} {\bibinfo {author} {\bibfnamefont {M.}~\bibnamefont
  {Cacciari}}, \bibinfo {author} {\bibfnamefont {G.~P.}\ \bibnamefont {Salam}},
  \ and\ \bibinfo {author} {\bibfnamefont {G.}~\bibnamefont {Soyez}},\ }\href
  {\doibase 10.1088/1126-6708/2008/04/063} {\bibfield  {journal} {\bibinfo
  {journal} {JHEP}\ }\textbf {\bibinfo {volume} {04}},\ \bibinfo {pages} {063}
  (\bibinfo {year} {2008})},\ \Eprint {http://arxiv.org/abs/0802.1189}
  {arXiv:0802.1189 [hep-ph]} \BibitemShut {NoStop}%
\bibitem [{\citenamefont {Cacciari}\ \emph {et~al.}(2012)\citenamefont
  {Cacciari}, \citenamefont {Salam},\ and\ \citenamefont
  {Soyez}}]{Cacciari:2011ma}%
  \BibitemOpen
  \bibfield  {author} {\bibinfo {author} {\bibfnamefont {M.}~\bibnamefont
  {Cacciari}}, \bibinfo {author} {\bibfnamefont {G.~P.}\ \bibnamefont {Salam}},
  \ and\ \bibinfo {author} {\bibfnamefont {G.}~\bibnamefont {Soyez}},\ }\href
  {\doibase 10.1140/epjc/s10052-012-1896-2} {\bibfield  {journal} {\bibinfo
  {journal} {Eur. Phys. J. C}\ }\textbf {\bibinfo {volume} {72}},\ \bibinfo
  {pages} {1896} (\bibinfo {year} {2012})},\ \Eprint
  {http://arxiv.org/abs/1111.6097} {arXiv:1111.6097 [hep-ph]} \BibitemShut
  {NoStop}%
\bibitem [{\citenamefont {Catani}\ \emph {et~al.}(2002)\citenamefont {Catani},
  \citenamefont {Dittmaier}, \citenamefont {Seymour},\ and\ \citenamefont
  {Trocsanyi}}]{Catani:2002hc}%
  \BibitemOpen
  \bibfield  {author} {\bibinfo {author} {\bibfnamefont {S.}~\bibnamefont
  {Catani}}, \bibinfo {author} {\bibfnamefont {S.}~\bibnamefont {Dittmaier}},
  \bibinfo {author} {\bibfnamefont {M.~H.}\ \bibnamefont {Seymour}}, \ and\
  \bibinfo {author} {\bibfnamefont {Z.}~\bibnamefont {Trocsanyi}},\ }\href
  {\doibase 10.1016/S0550-3213(02)00098-6} {\bibfield  {journal} {\bibinfo
  {journal} {Nucl. Phys. B}\ }\textbf {\bibinfo {volume} {627}},\ \bibinfo
  {pages} {189} (\bibinfo {year} {2002})},\ \Eprint
  {http://arxiv.org/abs/hep-ph/0201036} {arXiv:hep-ph/0201036} \BibitemShut
  {NoStop}%
\bibitem [{\citenamefont {Mehtar-Tani}\ and\ \citenamefont
  {Tywoniuk}(2017)}]{Mehtar-Tani:2016aco}%
  \BibitemOpen
  \bibfield  {author} {\bibinfo {author} {\bibfnamefont {Y.}~\bibnamefont
  {Mehtar-Tani}}\ and\ \bibinfo {author} {\bibfnamefont {K.}~\bibnamefont
  {Tywoniuk}},\ }\href {\doibase 10.1007/JHEP04(2017)125} {\bibfield  {journal}
  {\bibinfo  {journal} {JHEP}\ }\textbf {\bibinfo {volume} {04}},\ \bibinfo
  {pages} {125} (\bibinfo {year} {2017})},\ \Eprint
  {http://arxiv.org/abs/1610.08930} {arXiv:1610.08930 [hep-ph]} \BibitemShut
  {NoStop}%
\bibitem [{\citenamefont {Caucal}\ \emph {et~al.}(2022)\citenamefont {Caucal},
  \citenamefont {Soto-Ontoso},\ and\ \citenamefont {Takacs}}]{Caucal:2021cfb}%
  \BibitemOpen
  \bibfield  {author} {\bibinfo {author} {\bibfnamefont {P.}~\bibnamefont
  {Caucal}}, \bibinfo {author} {\bibfnamefont {A.}~\bibnamefont {Soto-Ontoso}},
  \ and\ \bibinfo {author} {\bibfnamefont {A.}~\bibnamefont {Takacs}},\ }\href
  {\doibase 10.1103/PhysRevD.105.114046} {\bibfield  {journal} {\bibinfo
  {journal} {Phys. Rev. D}\ }\textbf {\bibinfo {volume} {105}},\ \bibinfo
  {pages} {114046} (\bibinfo {year} {2022})},\ \Eprint
  {http://arxiv.org/abs/2111.14768} {arXiv:2111.14768 [hep-ph]} \BibitemShut
  {NoStop}%
\bibitem [{\citenamefont {Pablos}\ and\ \citenamefont
  {Soto-Ontoso}(2022)}]{Pablos:2022mrx}%
  \BibitemOpen
  \bibfield  {author} {\bibinfo {author} {\bibfnamefont {D.}~\bibnamefont
  {Pablos}}\ and\ \bibinfo {author} {\bibfnamefont {A.}~\bibnamefont
  {Soto-Ontoso}},\ }\href@noop {} {\  (\bibinfo {year} {2022})},\ \Eprint
  {http://arxiv.org/abs/2210.07901} {arXiv:2210.07901 [hep-ph]} \BibitemShut
  {NoStop}%
\bibitem [{\citenamefont {Blok}\ and\ \citenamefont
  {Tywoniuk}(2019)}]{Blok:2019uny}%
  \BibitemOpen
  \bibfield  {author} {\bibinfo {author} {\bibfnamefont {B.}~\bibnamefont
  {Blok}}\ and\ \bibinfo {author} {\bibfnamefont {K.}~\bibnamefont
  {Tywoniuk}},\ }\href {\doibase 10.1140/epjc/s10052-019-7061-4} {\bibfield
  {journal} {\bibinfo  {journal} {Eur. Phys. J. C}\ }\textbf {\bibinfo {volume}
  {79}},\ \bibinfo {pages} {560} (\bibinfo {year} {2019})},\ \Eprint
  {http://arxiv.org/abs/1901.07864} {arXiv:1901.07864 [hep-ph]} \BibitemShut
  {NoStop}%
\bibitem [{\citenamefont {Attems}\ \emph {et~al.}(2022)\citenamefont {Attems},
  \citenamefont {Brewer}, \citenamefont {Innocenti}, \citenamefont
  {Mazeliauskas}, \citenamefont {Park}, \citenamefont {van~der Schee},\ and\
  \citenamefont {Wiedemann}}]{Attems:2022otp}%
  \BibitemOpen
  \bibfield  {author} {\bibinfo {author} {\bibfnamefont {M.}~\bibnamefont
  {Attems}}, \bibinfo {author} {\bibfnamefont {J.}~\bibnamefont {Brewer}},
  \bibinfo {author} {\bibfnamefont {G.~M.}\ \bibnamefont {Innocenti}}, \bibinfo
  {author} {\bibfnamefont {A.}~\bibnamefont {Mazeliauskas}}, \bibinfo {author}
  {\bibfnamefont {S.}~\bibnamefont {Park}}, \bibinfo {author} {\bibfnamefont
  {W.}~\bibnamefont {van~der Schee}}, \ and\ \bibinfo {author} {\bibfnamefont
  {U.}~\bibnamefont {Wiedemann}},\ }\href@noop {} {\  (\bibinfo {year}
  {2022})},\ \Eprint {http://arxiv.org/abs/2209.13600} {arXiv:2209.13600
  [hep-ph]} \BibitemShut {NoStop}%
\bibitem [{\citenamefont {Baier}\ \emph
  {et~al.}(1997{\natexlab{a}})\citenamefont {Baier}, \citenamefont
  {Dokshitzer}, \citenamefont {Mueller}, \citenamefont {Peigne},\ and\
  \citenamefont {Schiff}}]{Baier:1996sk}%
  \BibitemOpen
  \bibfield  {author} {\bibinfo {author} {\bibfnamefont {R.}~\bibnamefont
  {Baier}}, \bibinfo {author} {\bibfnamefont {Y.~L.}\ \bibnamefont
  {Dokshitzer}}, \bibinfo {author} {\bibfnamefont {A.~H.}\ \bibnamefont
  {Mueller}}, \bibinfo {author} {\bibfnamefont {S.}~\bibnamefont {Peigne}}, \
  and\ \bibinfo {author} {\bibfnamefont {D.}~\bibnamefont {Schiff}},\ }\href
  {\doibase 10.1016/S0550-3213(96)00581-0} {\bibfield  {journal} {\bibinfo
  {journal} {Nucl. Phys. B}\ }\textbf {\bibinfo {volume} {484}},\ \bibinfo
  {pages} {265} (\bibinfo {year} {1997}{\natexlab{a}})},\ \Eprint
  {http://arxiv.org/abs/hep-ph/9608322} {arXiv:hep-ph/9608322} \BibitemShut
  {NoStop}%
\bibitem [{\citenamefont {Baier}\ \emph
  {et~al.}(1997{\natexlab{b}})\citenamefont {Baier}, \citenamefont
  {Dokshitzer}, \citenamefont {Mueller}, \citenamefont {Peigne},\ and\
  \citenamefont {Schiff}}]{Baier:1996kr}%
  \BibitemOpen
  \bibfield  {author} {\bibinfo {author} {\bibfnamefont {R.}~\bibnamefont
  {Baier}}, \bibinfo {author} {\bibfnamefont {Y.~L.}\ \bibnamefont
  {Dokshitzer}}, \bibinfo {author} {\bibfnamefont {A.~H.}\ \bibnamefont
  {Mueller}}, \bibinfo {author} {\bibfnamefont {S.}~\bibnamefont {Peigne}}, \
  and\ \bibinfo {author} {\bibfnamefont {D.}~\bibnamefont {Schiff}},\ }\href
  {\doibase 10.1016/S0550-3213(96)00553-6} {\bibfield  {journal} {\bibinfo
  {journal} {Nucl. Phys. B}\ }\textbf {\bibinfo {volume} {483}},\ \bibinfo
  {pages} {291} (\bibinfo {year} {1997}{\natexlab{b}})},\ \Eprint
  {http://arxiv.org/abs/hep-ph/9607355} {arXiv:hep-ph/9607355} \BibitemShut
  {NoStop}%
\bibitem [{\citenamefont {Wiedemann}(2000)}]{Wiedemann:2000za}%
  \BibitemOpen
  \bibfield  {author} {\bibinfo {author} {\bibfnamefont {U.~A.}\ \bibnamefont
  {Wiedemann}},\ }\href {\doibase 10.1016/S0550-3213(00)00457-0} {\bibfield
  {journal} {\bibinfo  {journal} {Nucl. Phys. B}\ }\textbf {\bibinfo {volume}
  {588}},\ \bibinfo {pages} {303} (\bibinfo {year} {2000})},\ \Eprint
  {http://arxiv.org/abs/hep-ph/0005129} {arXiv:hep-ph/0005129} \BibitemShut
  {NoStop}%
\bibitem [{\citenamefont {Gyulassy}\ \emph {et~al.}(2000)\citenamefont
  {Gyulassy}, \citenamefont {Levai},\ and\ \citenamefont
  {Vitev}}]{Gyulassy:2000fs}%
  \BibitemOpen
  \bibfield  {author} {\bibinfo {author} {\bibfnamefont {M.}~\bibnamefont
  {Gyulassy}}, \bibinfo {author} {\bibfnamefont {P.}~\bibnamefont {Levai}}, \
  and\ \bibinfo {author} {\bibfnamefont {I.}~\bibnamefont {Vitev}},\ }\href
  {\doibase 10.1103/PhysRevLett.85.5535} {\bibfield  {journal} {\bibinfo
  {journal} {Phys. Rev. Lett.}\ }\textbf {\bibinfo {volume} {85}},\ \bibinfo
  {pages} {5535} (\bibinfo {year} {2000})},\ \Eprint
  {http://arxiv.org/abs/nucl-th/0005032} {arXiv:nucl-th/0005032} \BibitemShut
  {NoStop}%
\bibitem [{\citenamefont {Armesto}\ \emph {et~al.}(2012)\citenamefont
  {Armesto}, \citenamefont {Ma}, \citenamefont {Mehtar-Tani}, \citenamefont
  {Salgado},\ and\ \citenamefont {Tywoniuk}}]{Armesto:2011ir}%
  \BibitemOpen
  \bibfield  {author} {\bibinfo {author} {\bibfnamefont {N.}~\bibnamefont
  {Armesto}}, \bibinfo {author} {\bibfnamefont {H.}~\bibnamefont {Ma}},
  \bibinfo {author} {\bibfnamefont {Y.}~\bibnamefont {Mehtar-Tani}}, \bibinfo
  {author} {\bibfnamefont {C.~A.}\ \bibnamefont {Salgado}}, \ and\ \bibinfo
  {author} {\bibfnamefont {K.}~\bibnamefont {Tywoniuk}},\ }\href {\doibase
  10.1007/JHEP01(2012)109} {\bibfield  {journal} {\bibinfo  {journal} {JHEP}\
  }\textbf {\bibinfo {volume} {01}},\ \bibinfo {pages} {109} (\bibinfo {year}
  {2012})},\ \Eprint {http://arxiv.org/abs/1110.4343} {arXiv:1110.4343
  [hep-ph]} \BibitemShut {NoStop}%
\bibitem [{\citenamefont {Caucal}\ \emph {et~al.}(2018)\citenamefont {Caucal},
  \citenamefont {Iancu}, \citenamefont {Mueller},\ and\ \citenamefont
  {Soyez}}]{Caucal:2018dla}%
  \BibitemOpen
  \bibfield  {author} {\bibinfo {author} {\bibfnamefont {P.}~\bibnamefont
  {Caucal}}, \bibinfo {author} {\bibfnamefont {E.}~\bibnamefont {Iancu}},
  \bibinfo {author} {\bibfnamefont {A.~H.}\ \bibnamefont {Mueller}}, \ and\
  \bibinfo {author} {\bibfnamefont {G.}~\bibnamefont {Soyez}},\ }\href
  {\doibase 10.1103/PhysRevLett.120.232001} {\bibfield  {journal} {\bibinfo
  {journal} {Phys. Rev. Lett.}\ }\textbf {\bibinfo {volume} {120}},\ \bibinfo
  {pages} {232001} (\bibinfo {year} {2018})},\ \Eprint
  {http://arxiv.org/abs/1801.09703} {arXiv:1801.09703 [hep-ph]} \BibitemShut
  {NoStop}%
\bibitem [{\citenamefont {Barata}\ \emph {et~al.}(2021)\citenamefont {Barata},
  \citenamefont {Mehtar-Tani}, \citenamefont {Soto-Ontoso},\ and\ \citenamefont
  {Tywoniuk}}]{Barata:2021wuf}%
  \BibitemOpen
  \bibfield  {author} {\bibinfo {author} {\bibfnamefont {J.~a.}\ \bibnamefont
  {Barata}}, \bibinfo {author} {\bibfnamefont {Y.}~\bibnamefont {Mehtar-Tani}},
  \bibinfo {author} {\bibfnamefont {A.}~\bibnamefont {Soto-Ontoso}}, \ and\
  \bibinfo {author} {\bibfnamefont {K.}~\bibnamefont {Tywoniuk}},\ }\href
  {\doibase 10.1007/JHEP09(2021)153} {\bibfield  {journal} {\bibinfo  {journal}
  {JHEP}\ }\textbf {\bibinfo {volume} {09}},\ \bibinfo {pages} {153} (\bibinfo
  {year} {2021})},\ \Eprint {http://arxiv.org/abs/2106.07402} {arXiv:2106.07402
  [hep-ph]} \BibitemShut {NoStop}%
\bibitem [{\citenamefont {Isaksen}\ \emph {et~al.}(2022)\citenamefont
  {Isaksen}, \citenamefont {Takacs},\ and\ \citenamefont
  {Tywoniuk}}]{Isaksen:2022pkj}%
  \BibitemOpen
  \bibfield  {author} {\bibinfo {author} {\bibfnamefont {J.~H.}\ \bibnamefont
  {Isaksen}}, \bibinfo {author} {\bibfnamefont {A.}~\bibnamefont {Takacs}}, \
  and\ \bibinfo {author} {\bibfnamefont {K.}~\bibnamefont {Tywoniuk}},\
  }\href@noop {} {\  (\bibinfo {year} {2022})},\ \Eprint
  {http://arxiv.org/abs/2206.02811} {arXiv:2206.02811 [hep-ph]} \BibitemShut
  {NoStop}%
\bibitem [{\citenamefont {Blok}(2021)}]{Blok:2020jgo}%
  \BibitemOpen
  \bibfield  {author} {\bibinfo {author} {\bibfnamefont {B.}~\bibnamefont
  {Blok}},\ }\href {\doibase 10.1140/epjc/s10052-021-09616-5} {\bibfield
  {journal} {\bibinfo  {journal} {Eur. Phys. J. C}\ }\textbf {\bibinfo {volume}
  {81}},\ \bibinfo {pages} {832} (\bibinfo {year} {2021})},\ \Eprint
  {http://arxiv.org/abs/2009.00465} {arXiv:2009.00465 [hep-ph]} \BibitemShut
  {NoStop}%
\bibitem [{\citenamefont {Feal}\ and\ \citenamefont
  {Vazquez}(2018)}]{Feal:2018sml}%
  \BibitemOpen
  \bibfield  {author} {\bibinfo {author} {\bibfnamefont {X.}~\bibnamefont
  {Feal}}\ and\ \bibinfo {author} {\bibfnamefont {R.}~\bibnamefont {Vazquez}},\
  }\href {\doibase 10.1103/PhysRevD.98.074029} {\bibfield  {journal} {\bibinfo
  {journal} {Phys. Rev. D}\ }\textbf {\bibinfo {volume} {98}},\ \bibinfo
  {pages} {074029} (\bibinfo {year} {2018})},\ \Eprint
  {http://arxiv.org/abs/1811.01591} {arXiv:1811.01591 [hep-ph]} \BibitemShut
  {NoStop}%
\bibitem [{\citenamefont {Andres}\ \emph {et~al.}(2020)\citenamefont {Andres},
  \citenamefont {Apolin\'ario},\ and\ \citenamefont
  {Dominguez}}]{Andres:2020vxs}%
  \BibitemOpen
  \bibfield  {author} {\bibinfo {author} {\bibfnamefont {C.}~\bibnamefont
  {Andres}}, \bibinfo {author} {\bibfnamefont {L.}~\bibnamefont
  {Apolin\'ario}}, \ and\ \bibinfo {author} {\bibfnamefont {F.}~\bibnamefont
  {Dominguez}},\ }\href {\doibase 10.1007/JHEP07(2020)114} {\bibfield
  {journal} {\bibinfo  {journal} {JHEP}\ }\textbf {\bibinfo {volume} {07}},\
  \bibinfo {pages} {114} (\bibinfo {year} {2020})},\ \Eprint
  {http://arxiv.org/abs/2002.01517} {arXiv:2002.01517 [hep-ph]} \BibitemShut
  {NoStop}%
\bibitem [{\citenamefont {Andres}\ \emph {et~al.}(2021)\citenamefont {Andres},
  \citenamefont {Dominguez},\ and\ \citenamefont
  {Gonzalez~Martinez}}]{Andres:2020kfg}%
  \BibitemOpen
  \bibfield  {author} {\bibinfo {author} {\bibfnamefont {C.}~\bibnamefont
  {Andres}}, \bibinfo {author} {\bibfnamefont {F.}~\bibnamefont {Dominguez}}, \
  and\ \bibinfo {author} {\bibfnamefont {M.}~\bibnamefont
  {Gonzalez~Martinez}},\ }\href {\doibase 10.1007/JHEP03(2021)102} {\bibfield
  {journal} {\bibinfo  {journal} {JHEP}\ }\textbf {\bibinfo {volume} {03}},\
  \bibinfo {pages} {102} (\bibinfo {year} {2021})},\ \Eprint
  {http://arxiv.org/abs/2011.06522} {arXiv:2011.06522 [hep-ph]} \BibitemShut
  {NoStop}%
\bibitem [{\citenamefont {Djordjevic}\ \emph {et~al.}(2005)\citenamefont
  {Djordjevic}, \citenamefont {Gyulassy},\ and\ \citenamefont
  {Wicks}}]{Djordjevic:2004nq}%
  \BibitemOpen
  \bibfield  {author} {\bibinfo {author} {\bibfnamefont {M.}~\bibnamefont
  {Djordjevic}}, \bibinfo {author} {\bibfnamefont {M.}~\bibnamefont
  {Gyulassy}}, \ and\ \bibinfo {author} {\bibfnamefont {S.}~\bibnamefont
  {Wicks}},\ }\href {\doibase 10.1103/PhysRevLett.94.112301} {\bibfield
  {journal} {\bibinfo  {journal} {Phys. Rev. Lett.}\ }\textbf {\bibinfo
  {volume} {94}},\ \bibinfo {pages} {112301} (\bibinfo {year} {2005})},\
  \Eprint {http://arxiv.org/abs/hep-ph/0410372} {arXiv:hep-ph/0410372}
  \BibitemShut {NoStop}%
\bibitem [{\citenamefont {Mustafa}(2005)}]{Mustafa:2004dr}%
  \BibitemOpen
  \bibfield  {author} {\bibinfo {author} {\bibfnamefont {M.~G.}\ \bibnamefont
  {Mustafa}},\ }\href {\doibase 10.1103/PhysRevC.72.014905} {\bibfield
  {journal} {\bibinfo  {journal} {Phys. Rev. C}\ }\textbf {\bibinfo {volume}
  {72}},\ \bibinfo {pages} {014905} (\bibinfo {year} {2005})},\ \Eprint
  {http://arxiv.org/abs/hep-ph/0412402} {arXiv:hep-ph/0412402} \BibitemShut
  {NoStop}%
\bibitem [{\citenamefont {Ke}\ \emph {et~al.}(2018)\citenamefont {Ke},
  \citenamefont {Xu},\ and\ \citenamefont {Bass}}]{Ke:2018tsh}%
  \BibitemOpen
  \bibfield  {author} {\bibinfo {author} {\bibfnamefont {W.}~\bibnamefont
  {Ke}}, \bibinfo {author} {\bibfnamefont {Y.}~\bibnamefont {Xu}}, \ and\
  \bibinfo {author} {\bibfnamefont {S.~A.}\ \bibnamefont {Bass}},\ }\href
  {\doibase 10.1103/PhysRevC.98.064901} {\bibfield  {journal} {\bibinfo
  {journal} {Phys. Rev. C}\ }\textbf {\bibinfo {volume} {98}},\ \bibinfo
  {pages} {064901} (\bibinfo {year} {2018})},\ \Eprint
  {http://arxiv.org/abs/1806.08848} {arXiv:1806.08848 [nucl-th]} \BibitemShut
  {NoStop}%
\bibitem [{\citenamefont {Herzog}\ \emph {et~al.}(2006)\citenamefont {Herzog},
  \citenamefont {Karch}, \citenamefont {Kovtun}, \citenamefont {Kozcaz},\ and\
  \citenamefont {Yaffe}}]{Herzog:2006gh}%
  \BibitemOpen
  \bibfield  {author} {\bibinfo {author} {\bibfnamefont {C.~P.}\ \bibnamefont
  {Herzog}}, \bibinfo {author} {\bibfnamefont {A.}~\bibnamefont {Karch}},
  \bibinfo {author} {\bibfnamefont {P.}~\bibnamefont {Kovtun}}, \bibinfo
  {author} {\bibfnamefont {C.}~\bibnamefont {Kozcaz}}, \ and\ \bibinfo {author}
  {\bibfnamefont {L.~G.}\ \bibnamefont {Yaffe}},\ }\href {\doibase
  10.1088/1126-6708/2006/07/013} {\bibfield  {journal} {\bibinfo  {journal}
  {JHEP}\ }\textbf {\bibinfo {volume} {07}},\ \bibinfo {pages} {013} (\bibinfo
  {year} {2006})},\ \Eprint {http://arxiv.org/abs/hep-th/0605158}
  {arXiv:hep-th/0605158} \BibitemShut {NoStop}%
\bibitem [{\citenamefont {Sievert}\ \emph {et~al.}(2019)\citenamefont
  {Sievert}, \citenamefont {Vitev},\ and\ \citenamefont
  {Yoon}}]{Sievert:2019cwq}%
  \BibitemOpen
  \bibfield  {author} {\bibinfo {author} {\bibfnamefont {M.~D.}\ \bibnamefont
  {Sievert}}, \bibinfo {author} {\bibfnamefont {I.}~\bibnamefont {Vitev}}, \
  and\ \bibinfo {author} {\bibfnamefont {B.}~\bibnamefont {Yoon}},\ }\href
  {\doibase 10.1016/j.physletb.2019.06.019} {\bibfield  {journal} {\bibinfo
  {journal} {Phys. Lett. B}\ }\textbf {\bibinfo {volume} {795}},\ \bibinfo
  {pages} {502} (\bibinfo {year} {2019})},\ \Eprint
  {http://arxiv.org/abs/1903.06170} {arXiv:1903.06170 [hep-ph]} \BibitemShut
  {NoStop}%
\bibitem [{\citenamefont {Blok}(2020)}]{Blok:2020kkn}%
  \BibitemOpen
  \bibfield  {author} {\bibinfo {author} {\bibfnamefont {B.}~\bibnamefont
  {Blok}},\ }\href {\doibase 10.1140/epjc/s10052-020-8324-9} {\bibfield
  {journal} {\bibinfo  {journal} {Eur. Phys. J. C}\ }\textbf {\bibinfo {volume}
  {80}},\ \bibinfo {pages} {729} (\bibinfo {year} {2020})},\ \Eprint
  {http://arxiv.org/abs/2002.11233} {arXiv:2002.11233 [hep-ph]} \BibitemShut
  {NoStop}%
\bibitem [{\citenamefont {Armesto}\ \emph {et~al.}(2005)\citenamefont
  {Armesto}, \citenamefont {Dainese}, \citenamefont {Salgado},\ and\
  \citenamefont {Wiedemann}}]{Armesto:2005iq}%
  \BibitemOpen
  \bibfield  {author} {\bibinfo {author} {\bibfnamefont {N.}~\bibnamefont
  {Armesto}}, \bibinfo {author} {\bibfnamefont {A.}~\bibnamefont {Dainese}},
  \bibinfo {author} {\bibfnamefont {C.~A.}\ \bibnamefont {Salgado}}, \ and\
  \bibinfo {author} {\bibfnamefont {U.~A.}\ \bibnamefont {Wiedemann}},\ }\href
  {\doibase 10.1103/PhysRevD.71.054027} {\bibfield  {journal} {\bibinfo
  {journal} {Phys. Rev. D}\ }\textbf {\bibinfo {volume} {71}},\ \bibinfo
  {pages} {054027} (\bibinfo {year} {2005})},\ \Eprint
  {http://arxiv.org/abs/hep-ph/0501225} {arXiv:hep-ph/0501225} \BibitemShut
  {NoStop}%
\bibitem [{\citenamefont {Huang}\ \emph {et~al.}(2013)\citenamefont {Huang},
  \citenamefont {Kang},\ and\ \citenamefont {Vitev}}]{Huang:2013vaa}%
  \BibitemOpen
  \bibfield  {author} {\bibinfo {author} {\bibfnamefont {J.}~\bibnamefont
  {Huang}}, \bibinfo {author} {\bibfnamefont {Z.-B.}\ \bibnamefont {Kang}}, \
  and\ \bibinfo {author} {\bibfnamefont {I.}~\bibnamefont {Vitev}},\ }\href
  {\doibase 10.1016/j.physletb.2013.08.009} {\bibfield  {journal} {\bibinfo
  {journal} {Phys. Lett. B}\ }\textbf {\bibinfo {volume} {726}},\ \bibinfo
  {pages} {251} (\bibinfo {year} {2013})},\ \Eprint
  {http://arxiv.org/abs/1306.0909} {arXiv:1306.0909 [hep-ph]} \BibitemShut
  {NoStop}%
\bibitem [{\citenamefont {Chatrchyan}\ \emph {et~al.}(2014)\citenamefont
  {Chatrchyan} \emph {et~al.}}]{CMS:2013qak}%
  \BibitemOpen
  \bibfield  {author} {\bibinfo {author} {\bibfnamefont {S.}~\bibnamefont
  {Chatrchyan}} \emph {et~al.} (\bibinfo {collaboration} {CMS}),\ }\href
  {\doibase 10.1103/PhysRevLett.113.132301} {\bibfield  {journal} {\bibinfo
  {journal} {Phys. Rev. Lett.}\ }\textbf {\bibinfo {volume} {113}},\ \bibinfo
  {pages} {132301} (\bibinfo {year} {2014})},\ \bibinfo {note} {[Erratum:
  Phys.Rev.Lett. 115, 029903 (2015)]},\ \Eprint
  {http://arxiv.org/abs/1312.4198} {arXiv:1312.4198 [nucl-ex]} \BibitemShut
  {NoStop}%
\bibitem [{\citenamefont {ATLAS}(2022)}]{ATLAS:2022fgb}%
  \BibitemOpen
  \bibfield  {author} {\bibinfo {author} {\bibnamefont {ATLAS}},\ }\href@noop
  {} {\  (\bibinfo {year} {2022})},\ \Eprint {http://arxiv.org/abs/2204.13530}
  {arXiv:2204.13530 [nucl-ex]} \BibitemShut {NoStop}%
\bibitem [{\citenamefont {CMS}(2022)}]{CMS:2022btc}%
  \BibitemOpen
  \bibfield  {author} {\bibinfo {author} {\bibnamefont {CMS}},\ }\href@noop {}
  {\  (\bibinfo {year} {2022})},\ \Eprint {http://arxiv.org/abs/2210.08547}
  {arXiv:2210.08547 [hep-ex]} \BibitemShut {NoStop}%
\bibitem [{\citenamefont {Ke}\ \emph {et~al.}(2021)\citenamefont {Ke},
  \citenamefont {Wang}, \citenamefont {Fan},\ and\ \citenamefont
  {Bass}}]{Ke:2020nsm}%
  \BibitemOpen
  \bibfield  {author} {\bibinfo {author} {\bibfnamefont {W.}~\bibnamefont
  {Ke}}, \bibinfo {author} {\bibfnamefont {X.-N.}\ \bibnamefont {Wang}},
  \bibinfo {author} {\bibfnamefont {W.}~\bibnamefont {Fan}}, \ and\ \bibinfo
  {author} {\bibfnamefont {S.~A.}\ \bibnamefont {Bass}},\ }\href {\doibase
  10.22323/1.387.0060} {\bibfield  {journal} {\bibinfo  {journal} {PoS}\
  }\textbf {\bibinfo {volume} {HardProbes2020}},\ \bibinfo {pages} {060}
  (\bibinfo {year} {2021})},\ \Eprint {http://arxiv.org/abs/2008.07622}
  {arXiv:2008.07622 [nucl-th]} \BibitemShut {NoStop}%
\bibitem [{\citenamefont {Li}\ and\ \citenamefont
  {Vitev}(2019{\natexlab{b}})}]{Li:2018xuv}%
  \BibitemOpen
  \bibfield  {author} {\bibinfo {author} {\bibfnamefont {H.~T.}\ \bibnamefont
  {Li}}\ and\ \bibinfo {author} {\bibfnamefont {I.}~\bibnamefont {Vitev}},\
  }\href {\doibase 10.1007/JHEP07(2019)148} {\bibfield  {journal} {\bibinfo
  {journal} {JHEP}\ }\textbf {\bibinfo {volume} {07}},\ \bibinfo {pages} {148}
  (\bibinfo {year} {2019}{\natexlab{b}})},\ \Eprint
  {http://arxiv.org/abs/1811.07905} {arXiv:1811.07905 [hep-ph]} \BibitemShut
  {NoStop}%
\bibitem [{\citenamefont {Acharya}\ \emph {et~al.}(2020)\citenamefont {Acharya}
  \emph {et~al.}}]{ALICE:2019qyj}%
  \BibitemOpen
  \bibfield  {author} {\bibinfo {author} {\bibfnamefont {S.}~\bibnamefont
  {Acharya}} \emph {et~al.} (\bibinfo {collaboration} {ALICE}),\ }\href
  {\doibase 10.1103/PhysRevC.101.034911} {\bibfield  {journal} {\bibinfo
  {journal} {Phys. Rev. C}\ }\textbf {\bibinfo {volume} {101}},\ \bibinfo
  {pages} {034911} (\bibinfo {year} {2020})},\ \Eprint
  {http://arxiv.org/abs/1909.09718} {arXiv:1909.09718 [nucl-ex]} \BibitemShut
  {NoStop}%
\bibitem [{\citenamefont {Berta}\ \emph {et~al.}(2019)\citenamefont {Berta},
  \citenamefont {Masetti}, \citenamefont {Miller},\ and\ \citenamefont
  {Spousta}}]{Berta:2019hnj}%
  \BibitemOpen
  \bibfield  {author} {\bibinfo {author} {\bibfnamefont {P.}~\bibnamefont
  {Berta}}, \bibinfo {author} {\bibfnamefont {L.}~\bibnamefont {Masetti}},
  \bibinfo {author} {\bibfnamefont {D.~W.}\ \bibnamefont {Miller}}, \ and\
  \bibinfo {author} {\bibfnamefont {M.}~\bibnamefont {Spousta}},\ }\href
  {\doibase 10.1007/JHEP08(2019)175} {\bibfield  {journal} {\bibinfo  {journal}
  {JHEP}\ }\textbf {\bibinfo {volume} {08}},\ \bibinfo {pages} {175} (\bibinfo
  {year} {2019})},\ \Eprint {http://arxiv.org/abs/1905.03470} {arXiv:1905.03470
  [hep-ph]} \BibitemShut {NoStop}%
\bibitem [{\citenamefont {Andres}\ \emph {et~al.}(2022)\citenamefont {Andres},
  \citenamefont {Dominguez}, \citenamefont {Kunnawalkam~Elayavalli},
  \citenamefont {Holguin}, \citenamefont {Marquet},\ and\ \citenamefont
  {Moult}}]{Andres:2022ovj}%
  \BibitemOpen
  \bibfield  {author} {\bibinfo {author} {\bibfnamefont {C.}~\bibnamefont
  {Andres}}, \bibinfo {author} {\bibfnamefont {F.}~\bibnamefont {Dominguez}},
  \bibinfo {author} {\bibfnamefont {R.}~\bibnamefont {Kunnawalkam~Elayavalli}},
  \bibinfo {author} {\bibfnamefont {J.}~\bibnamefont {Holguin}}, \bibinfo
  {author} {\bibfnamefont {C.}~\bibnamefont {Marquet}}, \ and\ \bibinfo
  {author} {\bibfnamefont {I.}~\bibnamefont {Moult}},\ }\href@noop {} {\
  (\bibinfo {year} {2022})},\ \Eprint {http://arxiv.org/abs/2209.11236}
  {arXiv:2209.11236 [hep-ph]} \BibitemShut {NoStop}%
\bibitem [{\citenamefont {Zigic}\ \emph {et~al.}(2019)\citenamefont {Zigic},
  \citenamefont {Salom}, \citenamefont {Auvinen}, \citenamefont {Djordjevic},\
  and\ \citenamefont {Djordjevic}}]{Zigic:2018smz}%
  \BibitemOpen
  \bibfield  {author} {\bibinfo {author} {\bibfnamefont {D.}~\bibnamefont
  {Zigic}}, \bibinfo {author} {\bibfnamefont {I.}~\bibnamefont {Salom}},
  \bibinfo {author} {\bibfnamefont {J.}~\bibnamefont {Auvinen}}, \bibinfo
  {author} {\bibfnamefont {M.}~\bibnamefont {Djordjevic}}, \ and\ \bibinfo
  {author} {\bibfnamefont {M.}~\bibnamefont {Djordjevic}},\ }\href {\doibase
  10.1088/1361-6471/ab2356} {\bibfield  {journal} {\bibinfo  {journal} {J.
  Phys. G}\ }\textbf {\bibinfo {volume} {46}},\ \bibinfo {pages} {085101}
  (\bibinfo {year} {2019})},\ \Eprint {http://arxiv.org/abs/1805.03494}
  {arXiv:1805.03494 [nucl-th]} \BibitemShut {NoStop}%
\bibitem [{\citenamefont {Fan}\ \emph {et~al.}(2022)\citenamefont {Fan} \emph
  {et~al.}}]{JETSCAPE:2022hcb}%
  \BibitemOpen
  \bibfield  {author} {\bibinfo {author} {\bibfnamefont {W.}~\bibnamefont
  {Fan}} \emph {et~al.} (\bibinfo {collaboration} {JETSCAPE}),\ }\href@noop {}
  {\  (\bibinfo {year} {2022})},\ \Eprint {http://arxiv.org/abs/2208.00983}
  {arXiv:2208.00983 [nucl-th]} \BibitemShut {NoStop}%
\end{thebibliography}%
\bibliographystyle{apsrev4-1}

\end{document}